\documentclass[10pt, reprint, twocolumn, nofootinbib, showkeys, aps, prd, colorlinks=true, linkcolor=blue, urlcolor=blue, citecolor=red, anchorcolor=blue]{revtex4-2}
\usepackage{amsmath}
\usepackage{amsfonts,txfonts,pxfonts}
\usepackage{amssymb}
\usepackage{orcidlink}
\usepackage[export]{adjustbox}
\usepackage{fix-cm}
\usepackage{adjustbox}
\usepackage{latexsym,amsmath,amssymb}
\RequirePackage{graphicx}
\RequirePackage{mathptmx}
\usepackage[arrowdel]{physics}
\usepackage{graphicx,epstopdf}
\usepackage{slashed}
\usepackage{bm}
\usepackage{cases}
\usepackage{color}
\usepackage{bm}
\usepackage{dcolumn}
\usepackage{lipsum}
\usepackage{hyperref}
\begin{document}
\title{Cosmological dynamics and structure formation in Tsallis-Cirto entropy-inspired modified gravity}
\author{Subhra Mondal \orcidlink{0009-0003-6469-6238}}
 \email{cosmology313@gmail.com}
\author{Amitava Choudhuri \orcidlink{0000-0001-9499-8585}}
 \email{amitava\_ch26@yahoo.com}
\affiliation{Department of Physics, \href{https://ror.org/05cyd8v32}{The University of Burdwan}, Golapbag, Purba Bardhaman - 713104, West Bengal, India.}
\hspace{1cm}
\begin{abstract}
We explore the cosmological dynamics and formation of structures within the Tsallis-Cirto modified gravity framework with a model parameter $\beta$ that is inspired by nonextensive statistics. By applying the gravity-thermodynamics conjecture, we modify the Friedmann equations and the evolution of the Hubble parameter. The impact of the Tsallis-Cirto entropy on different cosmographic parameters and the development of the linear matter overdensities have been analyzed by formulating perturbed field equations using the spherical collapse approach in a flat Friedmann-Lema\^{i}tre-Robertson-Walker background. We present a novel and established diagnostic approach to distinguish among distinct cosmological models compared to flat and non-flat $\Lambda$CDM scenarios, discovering that the Tsallis-Cirto entropy-inspired modified cosmology ($\beta\ne 1$) successfully withstands all tests, contradicting both the flat and non-flat $\Lambda$CDM models. This model also meets the conditions for the Universe to reach thermodynamic equilibrium in the far future. Additionally, we investigate the halo mass function and cluster number counts within this modified gravity framework. All findings are compared to the fiducial $\Lambda$CDM profile, indicating that the added entropic correction affects the history of expansion, the growth rate of structures, and the dark matter halo abundance. We find that the more massive structures are less abundant and develop at later times, which aligns with the hierarchical model of formation of large-scale structures.
\end{abstract}
\maketitle
\tableofcontents
\section{Introduction}
\par In modern physics, General Relativity (GR) together with Quantum Field Theory (QFT) serves as a fundamental pillar and has proven its remarkable success \cite{carroll2019spacetime}. The cosmology stemming from GR, i.e., the $\Lambda$CDM standard model, has achieved notable milestones in explaining several cosmological aspects ranging from the power spectrum and statistical properties of Cosmic Microwave Background Radiation (CMBR) anisotropies \cite{page2003first} to accelerated expansion of the Universe in the later stages \cite{perlmutter1999measurements,riess1998observational}. This model offers a clear and efficient way for describing the characteristics of large-scale cosmological structures \cite{bernardeau2002large,bull2016beyond}, at the same time maintaining consistency with the observed abundance of light elements like helium and hydrogen \cite{schramm1998big,steigman2007primordial,cyburt2016big,iocco2009primordial}. However, regardless of the simplicity of the model and its impressive achievements, the viability of the $\Lambda$CDM framework is currently questioned due to some theoretical and observational challenges \cite{abdullah2022coherent,anchordoqui2021dissecting,buchert2016observational,schmitz2022modern,di2021realm}. Consequently, alternative theories of gravity or cosmology have been speculated to better explain the observed phenomena.
\par There are several theoretical approaches to develop a theory beyond both GR and the $\Lambda$CDM paradigm. The first approach involves modifying the geometric framework of gravity, which is analyzed through extensions of the Einstein-Hilbert action, instead of adhering to Einstein’s GR formulation. This leads to a different set of models collectively referred to as modified gravity theories \cite{nojiri2017modified, capozziello2011extended}. Another strategic approach retains GR as the base theory of gravity while altering the matter sector. This method incorporates novel dynamical quantities, such as dark energy (DE)-like fluids and inflaton fields, which serve an important role in describing cosmic acceleration \cite{Olive1989inflation, guth1981inflationary,copeland2006dynamics,sahni2006reconstructing,li2011dark}. An alternative but equally significant perspective proposes a profound connection between gravity and thermodynamics \cite{padmanabhan2005gravity,jacobson1995thermodynamics,eling2006nonequilibrium,padmanabhan2006gravity,kothawala2007einstein,padmanabhan2012dialogue,padmanabhan2002classical,nojiri2025modified}. In this framework, one can use the first law of thermodynamics at the boundary to derive the fundamental field equations where the Universe is considered as a thermodynamic system bounded by the apparent horizon \cite{akbar2006friedmann,akbar2007thermodynamic,cai2007unified,cai2005first}. This formulation can be extended to numerous modified gravity scenarios alongside standard GR, provided the corresponding area laws are properly generalized \cite{paranjape2006thermodynamic,akbar2006friedmann,cai2010horizon,jamil2010thermodynamics}. It is not a novel assertion that horizon entropy gets special attention in the realm of entropic cosmology, stemming from the extension of black hole (BH) thermodynamics to cosmological scales, along with the holographic principle and the concept of entropic force. The natural outcome of the entropic force is the cosmic acceleration, which emerges from the inclusion of the Gibbons-Hawking-York (GHY) boundary term into the standard Einstein-Hilbert action \cite{easson2011entropic,easson2012entropic}. The updated action for the gravitational field along with matter fields over a region
$\mathcal{M}$ with boundary $\partial\mathcal{M}$ takes the form\footnote{Throughout this manuscript, the lowercase Latin indices \(a, b, \ldots, i, j, \ldots\) denote the values \(0, 1, 2, 3\) unless stated otherwise. The index \(0\) represents the temporal dimension, while the indices \(1, 2, \) and \(3\) correspond to the typical spatial dimensions. We follow the $(-,+,+,+)$ convention of spacetime signature. Standard notations are employed for the universal constants, such as the speed of light \(c\), gravitational constant \(G\), Planck constant \(\hbar\), and Boltzmann constant \(k_B\).}
\begin{equation}
I = \underbrace{\int_{\mathcal{M}} \left( \frac{c^4 \mathcal{R}}{16\pi G} + \mathcal{L}_{matter} \right) \sqrt{\mathrm{-g}} \, d^4x}_{\text{Einstein-Hilbert + Matter terms }} + \underbrace{\frac{c^4}{8\pi G} \oint_{\partial\mathcal{M}} \mathcal{K} \sqrt{{\mathfrak{h}}} \, d^3y}_{\text{Gibbons-Hawking-York term}}
\end{equation}
where $\mathcal{R}$ represents the Ricci scalar, $\mathcal{L}_{matter}$ signifies the matter Lagrangian, $\mathrm{g}$ refers to the determinant of the covariant components of the bulk metric tensor $g_{ab}$, $\mathcal{K}$ represents the trace of the extrinsic curvature of the boundary, and $\mathfrak{h}$ denotes the determinant of the induced metric $\mathfrak{h}_{ab}$ that resides on the boundary \cite{hawking1996gravitational,gibbons1977action,basilakos2012generalizing}. The GHY boundary term included in the updated action is crucial for describing the thermodynamic characteristics of the cosmological horizon and results in modified Einstein’s field equations \cite{easson2011entropic,easson2012entropic,mondal2024dynamics,mondal2025dynamics}. In this context, it is essential to emphasize that entropic cosmology is formulated within the extended GR framework in the FLRW background. This approach contrasts with  Verlinde’s entropic gravity \cite{verlinde2011origin} and Padmanabhan's emergent gravity \cite{padmanabhan2012emergence} concepts. Despite their different origins, both approaches consider gravity as an emergent phenomenon instead of a fundamental force, providing a radical departure from traditional geometric approaches.
\par Numerious studies have explored modifications to the Bekenstein-Hawking area law within the framework of higher-order curvature theories (see \cite{cai2009hawking,cai2008corrected,cai2008thermodynamics,tsallis2013black,tsallis2019black,d2024lagrangian,barrow2020area,saridakis2020modified,mondal2024dynamics,mondal2025dynamics,mondal2026cosmological,mondal2026gravity,basilakos2025modified,gohar2024generalized,nojiri2025modified,luciano2026new,ghaffari2019black,lymperis2021modified,sheykhi2010thermodynamics,sheykhi2011power,sheykhi2010entropic,cai2008corrected,tsallis2009introduction,tsallis1988possible,renyi1961measures,sharma1975new,kaniadakis2002statistical,kaniadakis2001non,barrow2020area} and references included). Two plausible modifications that incorporate quantum effects are due to power-law and logarithmic corrections. The power-law modifications emerge when there is an entanglement between quantum fields inside and outside the event horizon \cite{sheykhi2011power,das2008power,radicella2010generalized}. The logarithmic modifications originate from loop quantum gravity and are a result of thermal fluctuations at equilibrium as well as quantum fluctuations \cite{das2002general,ashtekar1998quantum,zhang2008black,banerjee2008quantum,sheykhi2010thermodynamics}.
\par An alternative way of correcting the area law of BHs stems from the consideration that the thermodynamics of $D$-dimensional non-standard systems cannot be connected to the additive Boltzmann-Gibbs entropy $S_{BG}$ (or the von Neumann entropy in the context of quantum mechanical systems), but rather to generalized non-additive entropies. The BG entropy fails to represent systems with divergent partition functions, such as gravitational systems \cite{tsallis2013black,lyra1998nonextensivity,tsallis1998role,mondal2024dynamics}. The traditional area law of BH, established by Bekenstein and Hawking, has a fundamental limitation since the BH entropy breaches thermodynamic extensivity, as highlighted in \cite{tsallis2013black}. In this regard, it should be noted that if a thermodynamic system is to be appropriately classified as $(D-1)$-dimensional, then one can consider the additive entropy $S_{BG}$ as its thermodynamic entropy. On the contrary, in the case of being $D$-dimensional, $S_{BG}$ cannot be deemed its thermodynamic entropy; instead, a non-additive entropic functional must play that role \cite{tsallis2013black,rani2022tsallis,mondal2024dynamics}. The violation of such a thermodynamic principle concerning the area law is somewhat neglected or not taken into consideration. Indeed, it was not anticipated to preserve the thermodynamic extensivity of such intricate systems at that moment. Nonetheless, there are several mathematical and scientific observations that suggest such a perspective might be anomalous in nature. To resolve the paradox of non-standard complex systems (like strongly entangled systems, BHs, and systems obeying the area law in general), it is essential to suggest a generalized entropy that has non-additive characteristics \cite{tsallis2013black,rani2022tsallis}, when standard additive BG-von Neumann entropy does not reflect proportionality to its volume. Thus, a non-additive generalization of the definition of the entropy is introduced in an effort to keep the entropic extensivity of a thermodynamic system. In this respect, Hanel and Thurner \cite{hanel2011comprehensive,hanel2011generalized} emphasized the non-additive generalized forms of entropy to preserve the extensivity of entropy by Khinchine axioms and surface-dominant statistics for complex systems. The Tsallis-Cirto (TC) entropy is a generalization of the BG entropy \cite{tsallis2013black,tsallis2019black,tsallis1988possible}, which was proposed as a resolution of the thermodynamic dilemma. In this context, Tsallis and Cirto \cite{tsallis2013black,tsallis2019black} presented a microscopic mathematical formulation of BH entropy that indicates nonextensive statistics. In this scenario, the area-entropy relation of BHs is modified as $\mathbb{S}_{hor}\propto A_{hor}^\beta$, where the exponent $\beta$ is the non-extensive TC parameter, which quantifies the degree of non-extensivity of the system. 
\par The TC entropy has been studied by many researchers for many years because of its possible application in many different fields. So far, it has shown remarkable results in a variety of complex systems such as BHs \cite{tsallis2013black,tsallis2019black}, dark matter (DM) \cite{guha2019model,mondal2024dynamics}, holographic DE \cite{Tavayef:2018xwx,saridakis2018holographic,nojiri2021different,pandey2022new}, neutrinos \cite{Luciano2021qgeneralized,kaniadakis1996generalized}, self-gravitating stellar systems \cite{plastino1993stellar,hamity1996generalized}, background radiation \cite{tsallis1995generalization}, thermodynamic gravity \cite{Sheykhi_Friedmannequations, geng2020modified,luciano2022baryogenesis}, polymer chains \cite{jizba2017tsallis}, and low-dimensional dissipative systems \cite{lyra1998nonextensivity}. Teimoori et al. \cite{teimoori2024inflation} have investigated a slow-roll type inflationary scenario by reconstructing a $f(\mathcal{R})$ gravity model equivalent to TC entropy-based cosmology. They studied the power spectra of scalar and tensor perturbations and derived the inflationary observables such as the tensor-to-scalar ratio and the scalar spectral index. Their results showed better observational agreement with the Planck $2018$ CMB data \cite{akrami2020planck} in the framework of TC entropy-based inflation. In the work reported in \cite{ghoshal2021constraints}, the authors studied the consequences of the cosmology based on the TC entropy on the process of the light elements' formation in the early Universe, and studied its viability as a useful modified cosmological model. It is important to mention that the TC cosmology addresses the discrepancies between the current limits on DM relic abundance and the latest high-energy neutrino observations from IceCube data \cite{jizba2022tsallis}. By introducing modifications to the Friedmann equations based on TC entropy, the authors have reconciled these differences while maintaining the TC scaling exponent close to $1.57$ \cite{jizba2022tsallis}.  Basilakos et. al. \cite{basilakos2024alleviating} proposed a method to resolve both $H_0$ and $\sigma_8$ tensions simultaneously by applying a thermodynamics-gravity hypothesis exploiting non-additive TC entropy instead of the conventional Hawking-Bekenstein formulation. They indicate that, for a specific choice of the TC exponent $(\lesssim 1)$, it is possible to derive an effective phantom DE equation of state that alleviates the $H_0$ tension, introduces an increased friction term in the equation governing matter-perturbation evolution, and results in a relatively low effective Newton's constant that resolves the $\sigma_8$ tension. In the realm of large-scale structure formation, it is demonstrated that perturbations evolve at a faster (slower) rate in a TC-modified Universe for $\beta > 1$ ($\beta < 1$) compared to the standard Friedmann case i.e., $\beta = 1$) \cite{sheykhi2022growth}. In \cite{amendola2010dark}, it has been asserted that the age problem of the Universe cannot be resolved within the standard cosmological framework without considering a cosmological constant or any form of DE. Subsequently, Sheykhi \cite{SHEYKHI2018118} presented a solution to the age problem as well as late-time acceleration of the Universe within the context of TC cosmology without incorporating any type of DE. Therefore, from the early to the late-time stages of our Universe, the nonextensive TC entropy-modified cosmology may serve as a potential alternative to the standard FLRW cosmological model.
\par Current observations of distant Type Ia Supernovae provide evidence that confirms the Universe consists of roughly 5\% visible baryonic matter, 25\% DM, and 70\% DE \cite{perlmutter1999measurements,de2000flat,yang2020evidence}. There is a significant amount of evidence in support of the existence of DM and DE in our Universe \cite{seljak2006cosmological}. Recently, DE has gained attention as it is supposed to be a mysterious element that exerts negative pressure, helping to accelerate the Universe's expansion due to its anti-gravity characteristics. Another enigmatic component, DM, serves several essential roles in the evolution of the Universe. First, it provides the required gravitational pull for the rotation of spiral galaxies and galaxy clusters. Second, it provides the fluctuations necessary to develop in the early Universe, which lead to the observed cosmological structures. From the period of matter domination until decoupling, DM perturbations can grow as DM does not interact via electromagnetic forces like baryonic matter, allowing it to collapse under its own gravitational instability; in contrast, baryonic matter fluctuations cannot grow as they are closely linked to photons through Thomson scattering. Once decoupling occurs, the baryons become detached from the pressure support exerted by the photon radiation and quickly fall into potential wells of the DM, while the baryonic perturbations catch up with the DM perturbations within a few expansion times after the decoupling epoch. The baryonic matter follows the DM distribution due to gravity. As a result, galaxy clusters are found within the potential of DM halos. Therefore, the arrangement of galaxy clusters provides valuable insights regarding the distribution of DM halos throughout the Universe. Without DM, galaxies would have formed much later than we observe today \cite{kolb2018early,farsi2022structure}.
\par The development of large-scale structures in the Universe is an intriguing as well as complex issue in cosmology. It explains how our Universe transitioned from a smooth and uniform state to the highly clustered configuration we see in the present day. During the inflationary phase, tiny quantum fluctuations in the scalar curvature acted as foundations for large-scale cosmological structures. As the Universe expanded rapidly during inflation, these tiny disturbances were magnified, resulting in gravitational instabilities that ultimately formed the galaxies and clusters we observe today. Essentially, the regions that collapsed in the early Universe acted as the initial cosmic seeds for density fluctuations from which the massive structures emerged in the later Universe \cite{kolb2018early,peebles2020principles,white1978core}. In this context, the top-hat spherical collapse (SC) model provides the most straightforward and appropriate (semi-)analytical method to analyze the evolution of overdensities of infalling masses into a gravitationally bound system and the formation of structures \cite{gunn1972infall,abramo2007structure}. The SC equations can indeed be obtained from GR, assuming shear does not play a significant role \cite{gaztanaga2001nonlinear}. In this classical top-hat model, a uniform and symmetric spherical perturbation is studied across the perturbed area within a homogeneous background Universe \cite{fernandes2012spherical}. The inherent symmetry of this model leads us to explore a spherical perturbation within a Friedmann-Lema\^{i}tre-Robertson-Walker (FLRW) Universe. Essentially, the growth of perturbations in a spherical region can be described using the same Friedmann equations, albeit with a different scale factor that governs the underlying theory of gravity \cite{planelles2015large}. The SC model highlights that primordial spherical overdense regions expand according to the Hubble flow during the early stages. The procedure of collapse due to gravitational instability is closely tied to the dynamics of the background Hubble flow during the early epoch \cite{naderi2015evolution}. When the relative overdensity of spherical regions in comparison to the background remains very small, linear perturbation theory is adequate for investigating their evolutionary dynamics. At a certain point in time, gravity starts to dominate and surpass the rate of expansion. Eventually, the overdense sphere attains its maximum size and entirely separates from the background expansion. The subsequent phase is characterized by the collapse of the spherical region driven by its own self-gravity.
\par In the study of structure formation through the evolution of matter density fluctuations, numerous works have been conducted within diverse cosmological frameworks. These frameworks encompass reconstructed DE models \cite{mukherjee2025spherical}, Rastall gravity \cite{ziaie2020structure}, Dvali, Gabadadze, and Porrati (DGP) braneworld cosmology \cite{mukherjee2020spherical}, generalized mass-to-horizon entropy-inspired modified gravity \cite{mondal2026cosmological}, mimetic gravity \cite{farsi2022structure}, $\Lambda$ viscous cold dark matter (CDM) Universe \cite{velten2014structure}, energy-momentum squared gravity \cite{farsi2023evolution}, and among others \cite{brax2012structure,koyama2006structure}. Article \cite{abramo2007structure} investigates non-linear structure formation in the presence of DE perturbations. In a previous study in Ref. \cite{mondal2024temporal}, we explored the cosmological evolution of density perturbations of the Bose–Einstein condensate DM, using the Gross–Pitaevskii–Poisson (GPP) system through Jeans' instability formalism.
\par In this manuscript, we explore the dynamics of our Universe and the large-scale structure formation within a modified cosmological scenario that is inspired by a nonextensive property of entropy of the horizon. We intend to identify a possible candidate that can differentiate this model from the conventional $\Lambda$CDM framework and other alternative gravity theories. We investigate the impact of TC-entropic modifications in both non-perturbative and perturbative regimes, specifically in the later stages of the Universe, concentrating on the linear growth of structures and the abundance of collapsed halos of DM. These characteristics reveal a unique difference from the previous research presented in this modified cosmology in Ref. \cite{sheykhi2022growth}, which examines linear density perturbations only during the matter-dominated era. One might envision our work as an extension of theirs. 
\par This article is structured as follows. In Section \ref{TC-modified cosmology}, we present a concise overview of nonextensive TC-modified cosmology and examine its dynamics. In Sec. \ref{Growth of matter spherical overdensities in TC-modified cosmology}, we investigate the linear theory of matter perturbations within the spherical top-hat collapse method of the spatially flat TC-modified cosmology. For the same cosmological framework, we analyze the DM halo mass function and number counts of the collapsed structures in Sec. \ref{Halo mass function and cluster number counts in TC-modified cosmology}, employing the Sheth-Mo-Tormen approach. Section \ref{Discussion and Conclusions} is reserved for conclusions.
\section{Tsallis-Cirto entropy-inspired modified cosmology}\label{TC-modified cosmology}
\par We begin with a $(1+3)$-dimensional isotropic and homogeneous FLRW Universe having line elements \cite{bak2000cosmic,cai2005first}
\begin{equation}\label{metric}
ds^{2} = h_{ab}\, dX^{a} dX^{b} + \tilde{r}^{\,2}\, d\Omega_{2}^{2} ,
\end{equation}
where \( X^{0} = ct,\; X^{1} = r \), and \( \tilde{r} = a(t)\, r \). The $2$-dimensional metric is expressed as \( h_{ab} = \mathrm{diag}(-c^2,\; a^{2}(t)/(1 - \kappa r^{2})) \)
where \( \kappa \, = +1,\, 0,\, \text{and}\, -1 \)
denotes the spatial curvature constant for closed, flat, and open Universes, respectively. The line element of the $2$-dimentional unit sphere is represented by \(d\Omega_{2}^{\,2} = d\theta^2 + \sin^2{\theta}\,d\phi^2\). The scale factor \(a \equiv a(t)\) describes the cosmic expansion of our Universe. We consider that our Universe consists of non-relativistic pressureless matter, including visible matter, CDM, and DE, while ignoring radiation at lower redshifts, and model it in the form of a perfect fluid. The covariant components of energy-momentum tensor reads~\cite{carroll2019spacetime,mondal2024dynamics,mondal2025dynamics}
\begin{equation}\label{EMT}
    \mathcal{T}_{ab}= \left(\rho + \frac{P}{c^2}\right)u_a u_b + P g_{ab}\,,
\end{equation}
where $\rho$ and $P$ denote the density and pressure of the fluid-like matter, respectively. The components of the four-velocity $1$-form of the fluid, denoted as $u_a$, obey the normalization condition $u_a u^b=-1$. In a curved manifold, the conservation of energy-momentum $\nabla_a\mathcal{T}^{ab}=0$ leads to the continuity equation \cite{carroll2019spacetime,mondal2024dynamics,mondal2025dynamics}
\begin{equation}\label{continuity_equation}
    \dot{\rho} + 3H\left(\rho + \frac{P}{c^2}\right)=0\,\,.
\end{equation} 
Moreover, we assume our Universe is confined by the dynamical apparent horizon, a marginally trapped physical boundary with zero expansion, defined by the relation \( h_{ab}\, \partial^{a}\tilde{r}\, \partial^{b}\tilde{r} = 0 \), which corresponds to a radius \cite{bak2000cosmic,cai2005first,akbar2007thermodynamic,sheykhi2007thermodynamical,mondal2024dynamics,mondal2025dynamics}
\begin{equation}\label{apparent_horizon}
    \tilde{r}_{hor} = \frac{c}{\sqrt{H^2 + \kappa c^2/a^2}}\,.
\end{equation}
The Hubble parameter is identified by \(H \overset{\text{def}}{=} \dot{a}(t)/a(t)\). Throughout this article, overdots signify derivatives with respect to the cosmic time $t$. For a dynamic FLRW horizon, the surface gravity is given by \cite{akbar2007thermodynamic}
\begin{equation}
\kappa_{SG} = \frac{1}{2\sqrt{-h}}\, \partial_a \left( \sqrt{-h}\, h^{ab}\, \partial_b \tilde r \right) = -\frac{c^{2}}{\tilde{r}_{hor}}
\left(1 - \frac{\dot{\tilde r}_{hor}}{2 H \tilde r_{hor}}\right)\, .
\end{equation}
The temperature related to the apparent horizon can be expressed as \cite{akbar2007thermodynamic}
\begin{equation}\label{temp}
T_{hor} = \frac{\hbar\, |\kappa_{SG}|}{2\pi\, k_{B}\, c} = \frac{\hbar c}{2\pi k_{B}\tilde r_{hor}}\,\left|
1 - \frac{\dot{\tilde r}_{hor}}{2H\tilde r_{hor}}\right|\,.
\end{equation}
We further consider the first law of thermodynamics on the boundary of the Universe \cite{akbar2007thermodynamic}
\begin{equation}\label{first_law}
    dE= T_{hor} d\mathbb{S}_{hor} + \mathcal{W} dV_{hor} , 
\end{equation}
where the work density $\mathcal{W}=-\frac{1}{2}\,\, trace \mathcal{T}$ refers to an invariant of energy-momentum tensor $\mathcal{T}$ \cite{hayward1998unified,hayward1999dynamic}. The $trace$ represents the two-dimensional trace evaluated perpendicular to the spheres of symmetry. Hence, for an expanding Universe under the perfect fluid assumption
\begin{equation}\label{work term}
    \mathcal{W}=-\frac{1}{2}(\mathcal{T}^{00}g_{00}+\mathcal{T}^{11}g_{11})=\frac{1}{2}(\rho c^2 -P) \,.
\end{equation}
Our Universe with a three-dimensional sphere of radius $\tilde{r}_{hor}$ with area $A_{hor}=4\pi \tilde{r}_{hor}^2$ and volume $V_{hor}=\frac{4}{3}\pi \tilde{r}_{hor}^3$, contains total energy $E=\rho c^2 V_{hor}$. The entropy associated with the apparent horizon \cite{tsallis2013black}
\begin{equation}\label{TC entropy}
    \mathbb{S}_{hor} = \frac{k_B}{4L_{Pl}^{2\beta}} A_{hor}^\beta\,,
\end{equation}
follows the TC modification. Here $L_{Pl}=\sqrt{\hbar G/c^3}$ denotes the Planck's length. The model indicator $\beta$ refers to the TC parameter (or the nonextensive parameter) quantifying the degree of nonextensivity. When $\beta=1$, the famous Bekenstein-Hawking area law \cite{bekenstein1973black} is recovered. It is important to note that the extensivity of the entropy of a general $(D+1)$-dimensional system can be restored through the relation $\beta=D/(D-1)$ for $D>1$ \cite{tsallis2013black}. Specifically, $(3+1)$-dimensional Universe with $D=3$, $\beta=3/2$ explains extensivity. 
\par Now, merging Eqs. (\ref{temp}), (\ref{work term}), (\ref{TC entropy}) into the first law of Thermodynamics (\ref{first_law}) we obtain the modified Friedmann equation in TC cosmology \cite{mondal2024dynamics}
\begin{equation}\label{mod_Friedmann_eqn}
    \left(H^2 + \frac{\kappa c^2}{a^2}\right)^{2-\beta}=\Gamma_{\beta}\rho = \Gamma_{\beta}(\rho_m + \rho_{\Lambda}) \,,
\end{equation}
where $\Gamma_{\beta} = \frac{(4-2\beta)(4\pi)^{2-\beta}c^{5-2\beta} L_{Pl}^{2\beta}}{3\hbar\beta}$ is a $\beta$-dependent constant. The positivity of $\Gamma_{\beta}$ restricts the maximum value of $\beta,\,\, \beta < 2$. Here, $\rho_m$ and $\rho_\Lambda$ are the densities of matter and DE, respectively. One can recover the standard Friedmann equation for $\beta \rightarrow{1}$.
We typically introduce the modified cosmological density parameters for matter, DE, and spatial curvature as follows:
\begin{equation}\label{density_param}
    \Omega^{mod}_m \overset{\text{def}}{=} \frac{\rho_m}{\rho^{mod}_{cr}}\,,\quad  \Omega^{mod}_\Lambda \overset{\text{def}}{=} \frac{\rho_\Lambda}{\rho^{mod}_{cr}}\,,\quad \Omega^{mod}_\kappa \overset{\text{def}}{=} \frac{\kappa c^2}{a^2 H^2}\,,
\end{equation}
where $\rho^{mod}_{cr} = H^{4-2\beta}/\Gamma_{\beta}$ denotes the critical density in the TC-modified Universe. Thus, the first Friedmann equation (\ref{mod_Friedmann_eqn}) can be written by exploiting cosmological density parameters as
\begin{equation}
    \Omega^{mod}_m + \Omega^{mod}_{\Lambda} = (1+\Omega^{mod}_\kappa)^{2-\beta}\, .
\end{equation}
For a flat ($\kappa = 0$) TC Universe, the above relation reads
\begin{equation}
    \Omega^{mod}_m + \Omega^{mod}_{\Lambda} = 1\, .
\end{equation}
\begin{figure}[htbp!]
    \centering
    \includegraphics[width=8cm,height=6cm]{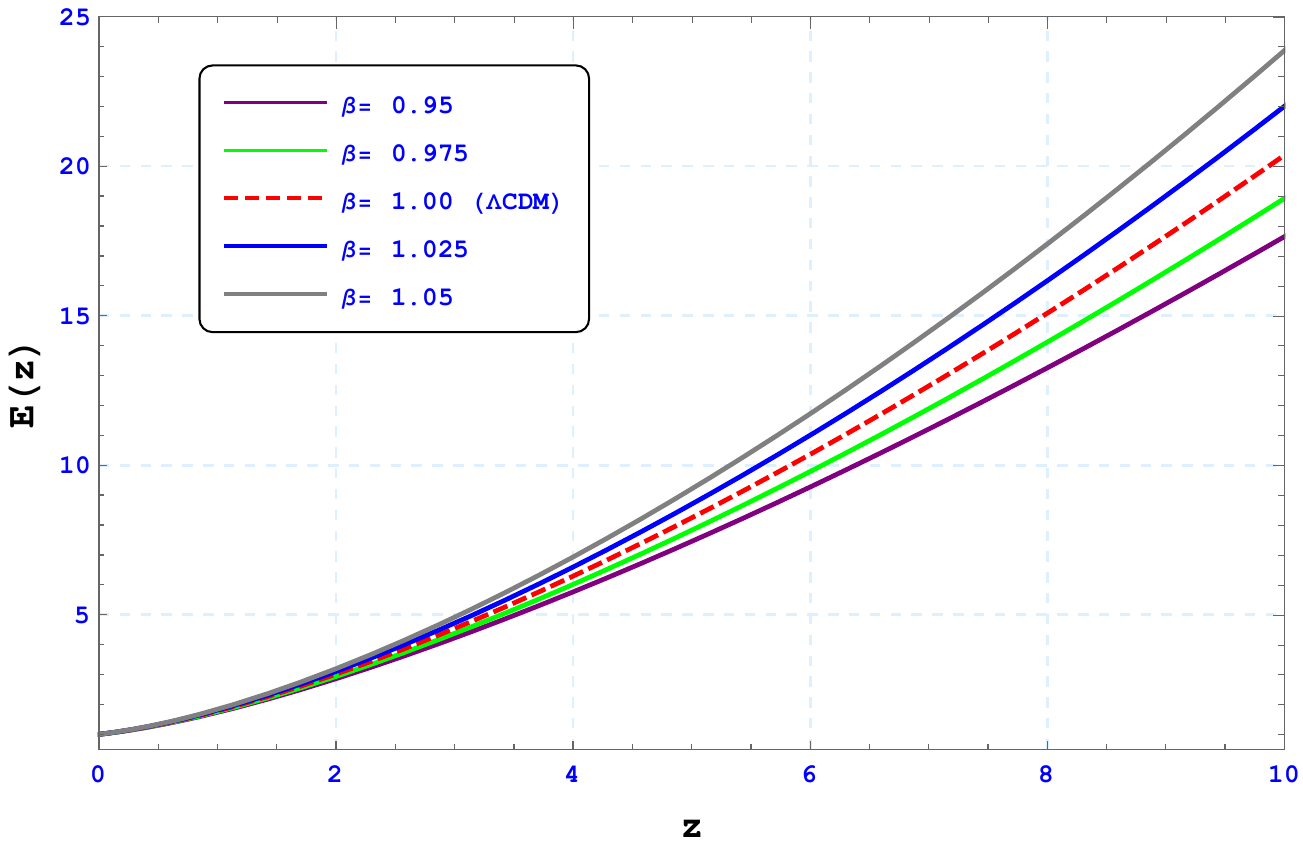}
    \caption{\small(Color online) Plot for normalized Hubble parameter $E(z)$ versus redshift $z$ with different values of the TC model parameter $\beta$}
    \label{ETC.pdf}
\end{figure}
We assume that pressureless matter and DE are fundamentally non-interacting; therefore, they adhere to distinct conservation equations as follows:
\begin{subequations}\label{contituity_eqn_seperate}
    \begin{align}
       \dot{\rho}_m + 3\frac{\dot{a}}{a}\rho_m &= 0 \,,\\
       \dot{\rho}_\Lambda &= 0 \,,  
    \end{align}
\end{subequations}
as derived from Eq. (\ref{continuity_equation}). The subsequent solutions for Eq. (\ref{contituity_eqn_seperate}) can be derived as
\begin{subequations}\label{solutions_matter_DE}
    \begin{align}
       \rho_m &= \rho_{m,0}a^{-3} =\rho_{m,0}(1+z)^{3} \,,\\
       \rho_\Lambda &= \rho_{\Lambda,0} \,.  
    \end{align}
\end{subequations}
Here, $\rho_{m,0}$ and $\rho_{\Lambda,0}$ are the current ($a=1$) densities of matter and DE sectors, respectively. We use the redshift-scale factor relation $z = \frac{1}{a}-1$, where the present-day scale factor is considered $a_0 = 1$. By inserting (\ref{solutions_matter_DE}) into Eq. (\ref{mod_Friedmann_eqn}) under a flat Universe scenario, we find the following Hubble parameter $H(z)$ and normalized Hubble parameter $E(z)$ in the TC-modified cosmology:
\begin{subequations}
    \begin{align}
       H(z) &= \Gamma_{\beta}^\frac{1}{4-2\beta}[\rho_{m} + \rho_{\Lambda}]^\frac{1}{4-2\beta} = \Gamma_{\beta}^\frac{1}{4-2\beta}[\rho_{m,0}(1+z)^{3} + \rho_{\Lambda,0}]^\frac{1}{4-2\beta} \,,\label{Hubble}\\
       E(z) &\overset{\text{def}}{=}  \frac{H(z)}{H_0} =  [\Omega_{m,0}(1+z)^{3}+\Omega_{\Lambda,0}]^\frac{1}{4-2\beta} \label{Normalized Hubble}\,. 
    \end{align}
\end{subequations}
In this context, we consider the Hubble parameter $H_0 = 100\,\mathrm{h} \text{ km s}^{-1} \text{ Mpc}^{-1}$ with $\mathrm{h} = 0.6766 \pm 0.0042$, the density parameters of matter and DE $\Omega_{m,0} = \frac{\rho_m}{\rho_{cr}} = 0.3111 \pm 0.0056$ and $\Omega_{\Lambda,0} = \frac{\rho_\Lambda}{\rho_{cr}} = 0.6889 \pm 0.0056$ \cite{aghanim2020planck,aghanim2021erratum}, respectively, at present ($z=0$), and $\rho_{cr}=\rho^{mod}_{cr}\big\rfloor_{\beta = 1}$ refers to the critical density in standard $\Lambda$CDM cosmology. At present, $E(z=0) = 1$. The evolution of the normalized Hubble parameter with redshift for different values of the model parameter $\beta$ is shown in Fig. \ref{ETC.pdf}. In the TC-modified cosmology, the normalized Hubble parameter increases as the model parameter $\beta$ changes from lower to higher values. Furthermore, for $\beta<1$ ($\beta>1$), the normalized Hubble parameter exhibits a gentler (steeper) slope, indicating that the expansion rate of this model Universe slows down (speeds up) compared with the standard $\Lambda$CDM framework.
\begin{figure}[htbp!]
    \centering
    \includegraphics[width=8cm,height=6cm]{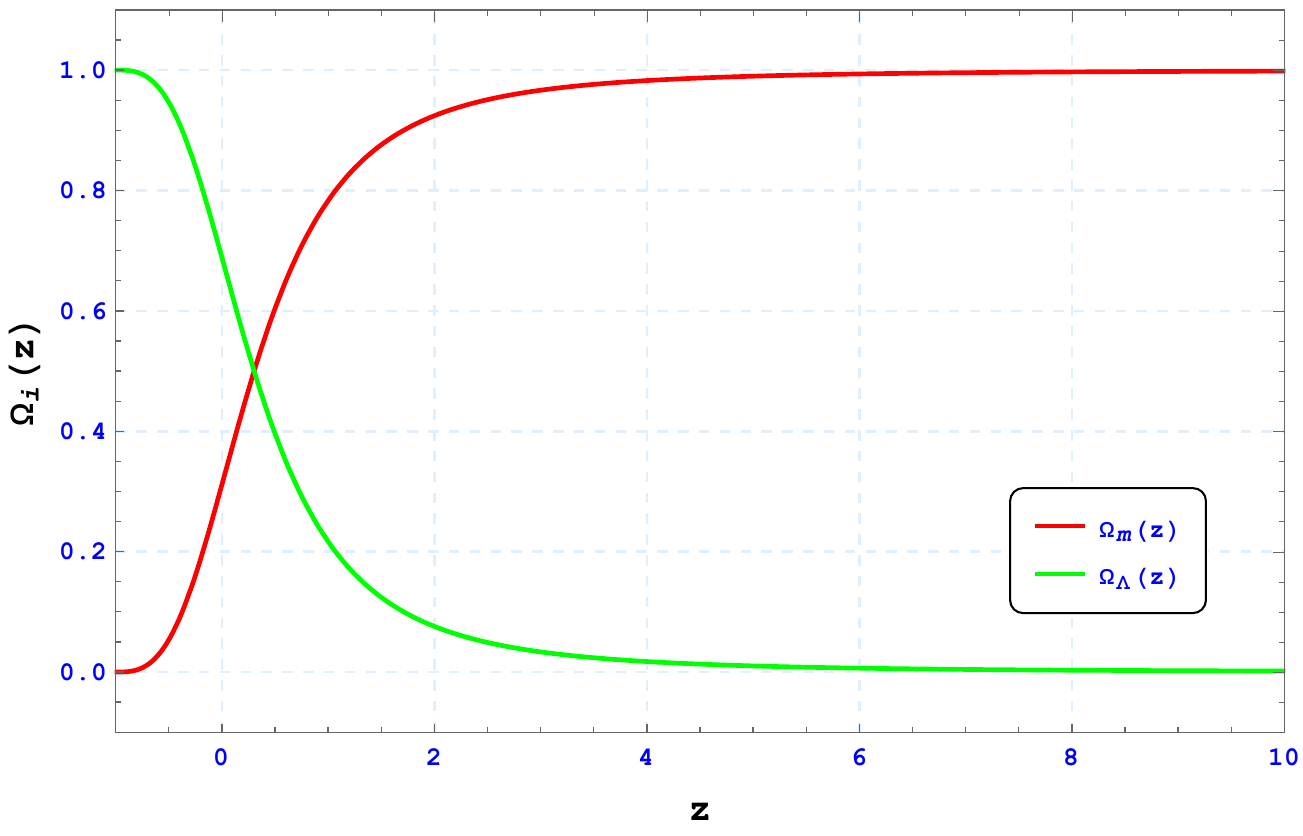}
    \caption{\small(Color online) Plot for cosmological density parameters $\Omega_i(z)\, (i\equiv m, \Lambda)$ versus redshift $z$}
    \label{OMEGA_iTC.pdf}
\end{figure}
\par Based on Eqs. (\ref{density_param}) and (\ref{Hubble}), the redshift evolution of matter and DE density parameters $\Omega^{mod}_m(z)$ and $\Omega^{mod}_{\Lambda}(z)$ in the TC cosmology can be represented as follows:
\begin{subequations}\label{density_parameters}
    \begin{align}
       \Omega^{mod}_m(z) &= \frac{\rho_m}{\rho_m + \rho_\Lambda} = \frac{\Omega_{m,0}(1+z)^{3}}{\Omega_{m,0}(1+z)^{3}+\Omega_{\Lambda,0}} = \Omega_m(z) \,,\label{matter_density_parameter}\\
       \Omega^{mod}_{\Lambda}(z) &= \frac{\rho_\Lambda}{\rho_m + \rho_\Lambda} = \frac{\Omega_{\Lambda,0}}{\Omega_{m,0}(1+z)^{3}+\Omega_{\Lambda,0}} = \Omega_{\Lambda}(z)\,\label{DE_density_parameter},  
    \end{align}
\end{subequations}
which do not depend on model parameters and show similarity with the standard flat $\Lambda$CDM cosmology satisfying the relation $\Omega_m + \Omega_\Lambda = 1$. The cosmological density parameters presented in Eq. (\ref{density_parameters}) are depicted in Fig. \ref{OMEGA_iTC.pdf}. The matter-DE equality epoch is pointed out at a redshift $z_{eq} = 0.303423$ akin to the $\Lambda$CDM model. We find $\Omega_\Lambda \to 1$ and $\Omega_m \to 0$ as $z\to -1$, suggesting a total DE-dominated Universe in the distant future, even though matter was predominant in the early Universe.
\begin{figure}[htbp!]
    \centering
    \includegraphics[width=8cm,height=6cm]{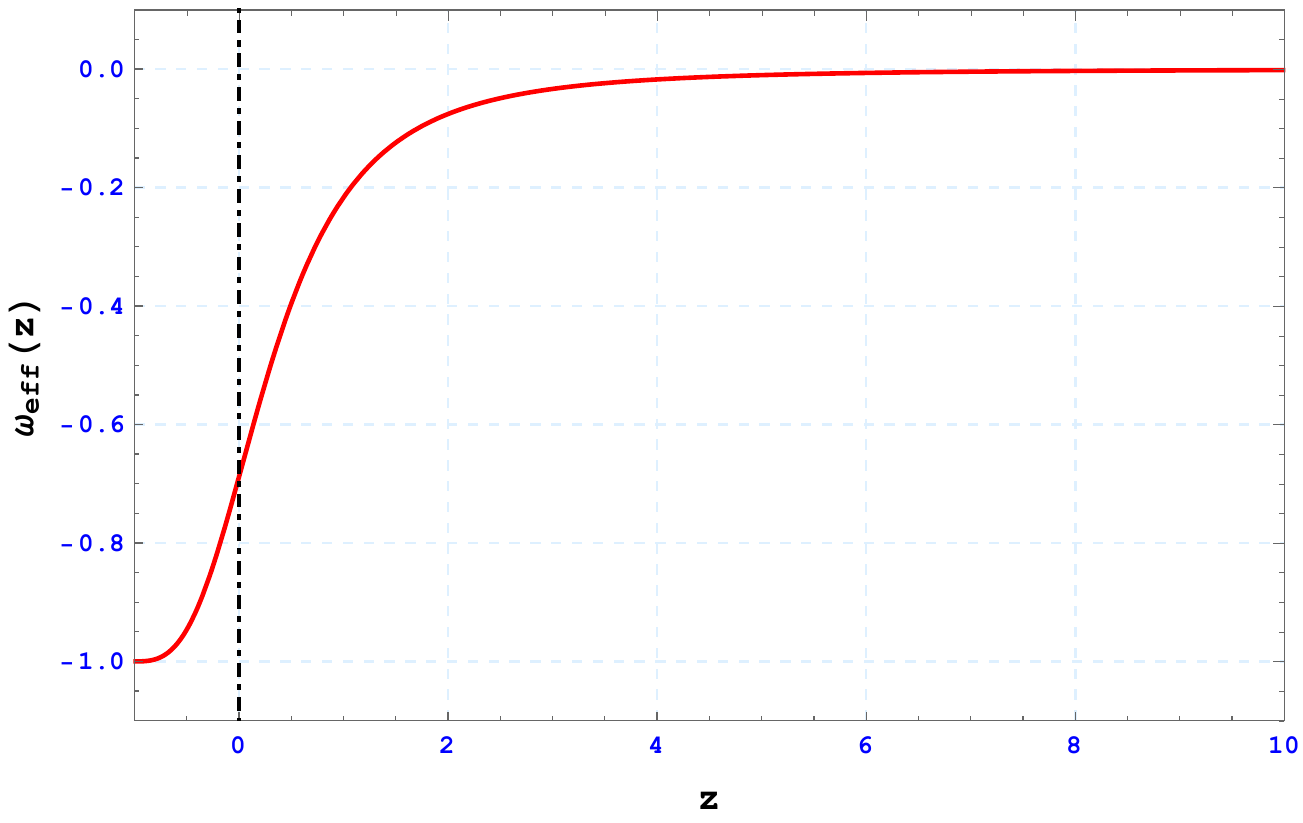}
    \caption{\small(Color online) Plot for effective equation of state parameter $w_{eff}$ versus redshift $z$}
    \label{omega_effTC.pdf}
\end{figure}
\par The effective equation of state $\omega_{\, eff}(z)$ parameter encapsulates the collective dynamics of the cosmological components that drive the expansion. It provides a straightforward and intuitive approach to monitor deviations from matter domination to DE domination. Treating the combination of pressureless matter and DE as a single fluid, one can obtain
\begin{equation}
    \omega_{\,eff}(z) \overset{\text{def}}{=} \frac{P_m + P_\Lambda}{c^2(\rho_m + \rho_\Lambda)} = -\frac{\Omega_{\Lambda,0}}{\Omega_{m,0}(1+z)^3 + \Omega_{\Lambda,0}}\, ,
\end{equation}
which remains independent of the model parameter and precisely aligns with the flat $\Lambda$CDM model. The redshift evolution of $\omega_{\, eff}(z)$ is presented in Fig. \ref{omega_effTC.pdf}. We find that $\omega_{\, eff} = -1$ indicates a transition to the de-Sitter phase in the distant future as $z\to -1$. Further, we get $\omega_{\, eff} =\Omega_{\Lambda,0} = -0.6889$ at present $z\to 0$, whereas in the very early Universe $z\to \infty$, we have $\omega_{\, eff} = 0$. This implies that the Universe undergoes early deceleration, followed by acceleration in a later epoch.
\par Cosmography \cite{visser2005cosmography,visser2004jerk} serves as a widely used geometric-based technique to examine the deviations of various modified cosmological models from the conventional $\Lambda$CDM profile. The cosmological scale factor for the FLRW Universe can be expanded in a Taylor series centred around the present epoch $t_0$ as $a(t)=\sum_{i=0}^{\infty}\frac{a^{(i)}(t_0)}{i!}(t-t_0)^i$, where $a^{(i)}(t_0)$ denotes the $i^{\text{th}}$ derivative of the scale factor $a(t)$ when $t=t_0$. The coefficients of the Taylor expansion correspond to the geometric quantities $H = \frac{\dot{a}}{a}$, $q = -\frac{\overset{\cdot\cdot}{a}}{aH^2}$, $j = \frac{\overset{\cdot\cdot\cdot}{a}}{aH^3}$, $s = \frac{\overset{\cdot\cdot\cdot\cdot}{a}}{aH^4}$, $\ell = \frac{\overset{\cdot\cdot\cdot\cdot\cdot}{a}}{aH^5}$, $m = \frac{\overset{\cdot\cdot\cdot\cdot\cdot\cdot}{a}}{aH^6}$, and so on, are referred to as the \textit{cosmographic parameters}. The newly introduced quantities, namely $q$, $j$, $s$, $\ell$, and $m$ are identified as the \textit{deceleration}, \textit{jerk}, \textit{snap}, \textit{lerk}, and \textit{$m$}-parameters \cite{pan2018astronomical,mondal2026cosmological}. Before the discovery of cosmic acceleration, observations centered primarily on $H$. As $H$ evolves with time, the next higher-order derivative of the scale factor, which is $q$, reflects that evolution. Currently, we find that $q$ itself changes over time, making the next derivative in the sequence, $j$, the natural choice for researchers to investigate \cite{mukherjee2016parametric}. This series can also be further extended to include additional kinematic quantities for a more comprehensive understanding.
\par
\begin{figure}[htbp!]
    \centering
    \includegraphics[width=8cm,height=6cm]{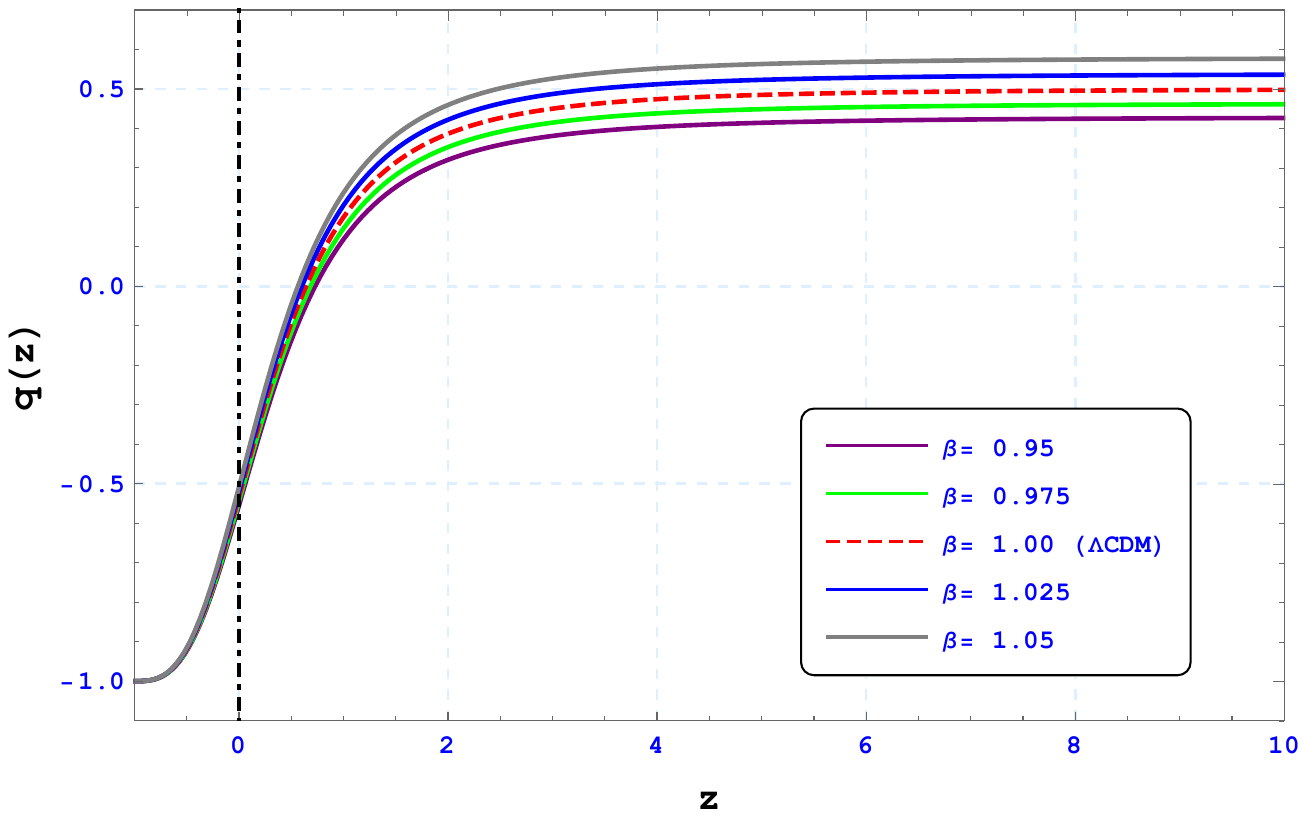}
    \caption{\small(Color online) Plot for deceleration paramater $q$ versus redshift $z$ with different values of TC model parameter $\beta$}
    \label{qTC.pdf}
\end{figure}

 The deceleration parameter $q$ serves as a kinematic indicator of cosmic expansion by measuring whether the Universe is expanding at an accelerating or decelerating rate. Its progression directly tracks the shift between these two states and reflects the impact of the main energy components influencing the expansion. Consequently, it offers a crucial method to investigate DE dynamics or modifications in cosmology and deviations from the conventional cosmological model. The deceleration parameter $q(z)$ within the framework of TC-modified cosmology can be defined in terms of redshift as follows:
\begin{align}\label{deceleration_param}
       q(z) &\overset{\text{def}}{=} -\frac{\overset{\cdot\cdot}{a}}{a H^2} = -\left(1+\frac{\dot{H}}{H}\right) = -1 + \frac{(1+z)}{E(z)}\frac{dE(z)}{dz}\nonumber\\[6pt]
       &= -1 + \left(\frac{3}{4-2\beta}\right)\frac{\Omega_{m,0}(1+z)^{3}}{\Omega_{m,0}(1+z)^{3}+\Omega_{\Lambda,0}} \, .
\end{align}
The transition redshift $z_{\mathrm{tr}}$ designates the moment in cosmic history when the expansion rate changes from deceleration ($q>0$) to acceleration ($q<0$), defined by the condition $q(z=z_{\, tr})=0$. This leads to the expression:
\begin{equation}
z_{\mathrm{tr}} =\left[\left(\frac{4-2\beta}{2\beta-1}\right)\,
\frac{\Omega_{\Lambda,0}}{\Omega_{m,0}}\right]^{\tfrac{1}{3}}
-1 \,.
\end{equation}
Figure \ref{qTC.pdf} effectively illustrates the transition from the deceleration phase to the acceleration phase for various values of $\beta$ at a redshift around $z_{\mathrm{tr}}\approx 0.64$, consistent with observational data \cite{myrzakulov2024linear,roman2019constraints,al2018observational}. It is also noted that $z_{\mathrm{tr}}$ is influenced by the model parameter $\beta$. Specifically, we find that $z_{\mathrm{tr}}$ decreases (increases) for $\beta > 1$ ($\beta < 1$), indicating that the onset of cosmic acceleration occurs later (earlier) in TC-modified cosmology compared to the standard $\Lambda$CDM model. Additionally, we observe that the deceleration parameter increases with rising values of $\beta$ during the early epoch. As $z \to -1$, the modified model approaches the $\Lambda$CDM model. As we see, $q \to -1$ when $z \to -1$; thus, the requirement for the Universe to reach thermodynamic equilibrium in the far future remains intact \cite{del2012three}.
\par
\begin{figure}[htbp!]
    \centering
    \includegraphics[width=8cm,height=6cm]{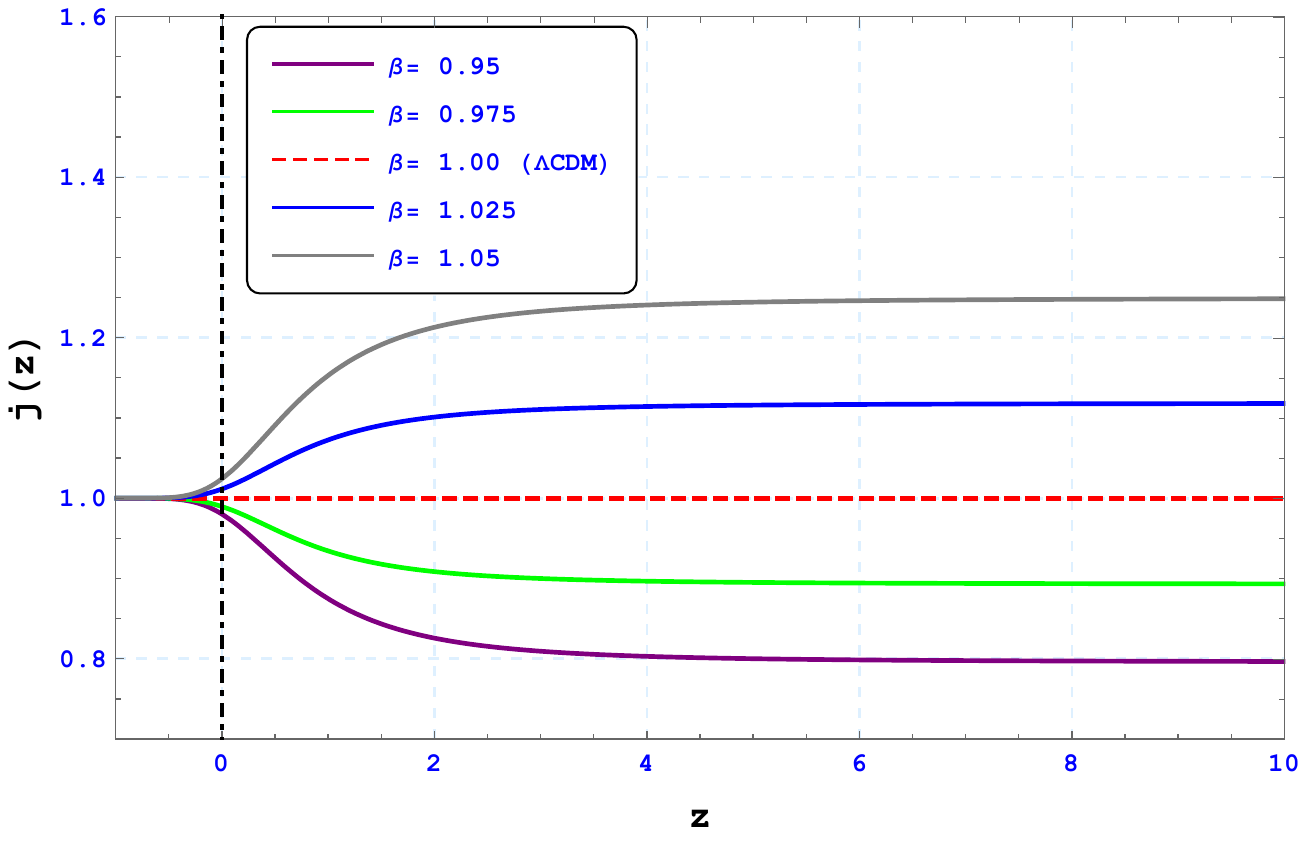}
    \caption{\small(Color online) Plot for jerk parameter $j(z)$ versus redshift $z$ with different values of TC model parameter $\beta$}
    \label{jTC.pdf}
\end{figure}

 The jerk parameter offers a higher-order kinematical viewpoint on the expansion of the Universe, enhancing the insights provided by the Hubble and deceleration parameters. While the second derivative of the scale factor, i.e., the deceleration parameter, governs the transition from deceleration to acceleration, the jerk parameter reveals the evolution of acceleration during this shift. The jerk parameter \( j(z) \) in the TC-modified cosmology can be written as
\begin{align}\label{jerk_parameter}
j(z) 
&\overset{\text{def}}{=} 
   \frac{\overset{\cdot\cdot\cdot}{a}}{a H^3} 
   = q(z)\,[2q(z)+1] + (1+z)\frac{dq(z)}{dz}
  \nonumber \\[6pt]
&= 
\frac{1}{
   \left[\Omega_{m,0}(1+z)^3 + \Omega_{\Lambda,0}\right]^2
   }
   \nonumber \\[4pt]
&\qquad\quad \times
\Bigg[
      \frac{2(2\beta-1)(\beta+1)}{(4-2\beta)^2}\,
      \Omega_{m,0}^2 (1+z)^6
      \nonumber \\[4pt]
&\qquad\quad
      + 2\,\Omega_{m,0}\,\Omega_{\Lambda,0} (1+z)^3 + \Omega_{\Lambda,0}^2
\Bigg]\,.
\end{align}
At the transition time, the relation
\begin{equation}
j(z_{\mathrm{tr}}) = (1+z_{\mathrm{tr}})\frac{dq}{dz}\Bigg\rfloor_{z=z_{\mathrm{tr}}}
\end{equation} 
illustrates that the jerk parameter effectively captures the smoothness of the transition from deceleration to acceleration. The redshift evolution of the jerk parameter is illustrated for various values of the model parameter \( \beta \) in Fig. \ref{jTC.pdf}. We note $j(z)$ remains constant at \( j(z) \big\rfloor_{\beta = 1} = 1 \) across all redshifts in the conventional \( \Lambda \)CDM model. Any deviation from this value indicates a distinct sign of physics beyond the traditional standard cosmological model. It is to be noted that the jerk parameter rises with increasing \( \beta \), and as \( z \to -1 \), the modified model aligns with the \( \Lambda \)CDM model. Interestingly, at \( z = 0 \), the jerk parameter diverges from unity, signifying a clear deviation from the predictions of \( \Lambda \)CDM at the present epoch. This observation highlights that the modification due to nonextensivity could impact late-time acceleration in the TC-modified cosmology.
\par
\begin{figure}[htbp!]
    \centering
    \includegraphics[width=8cm,height=6cm]{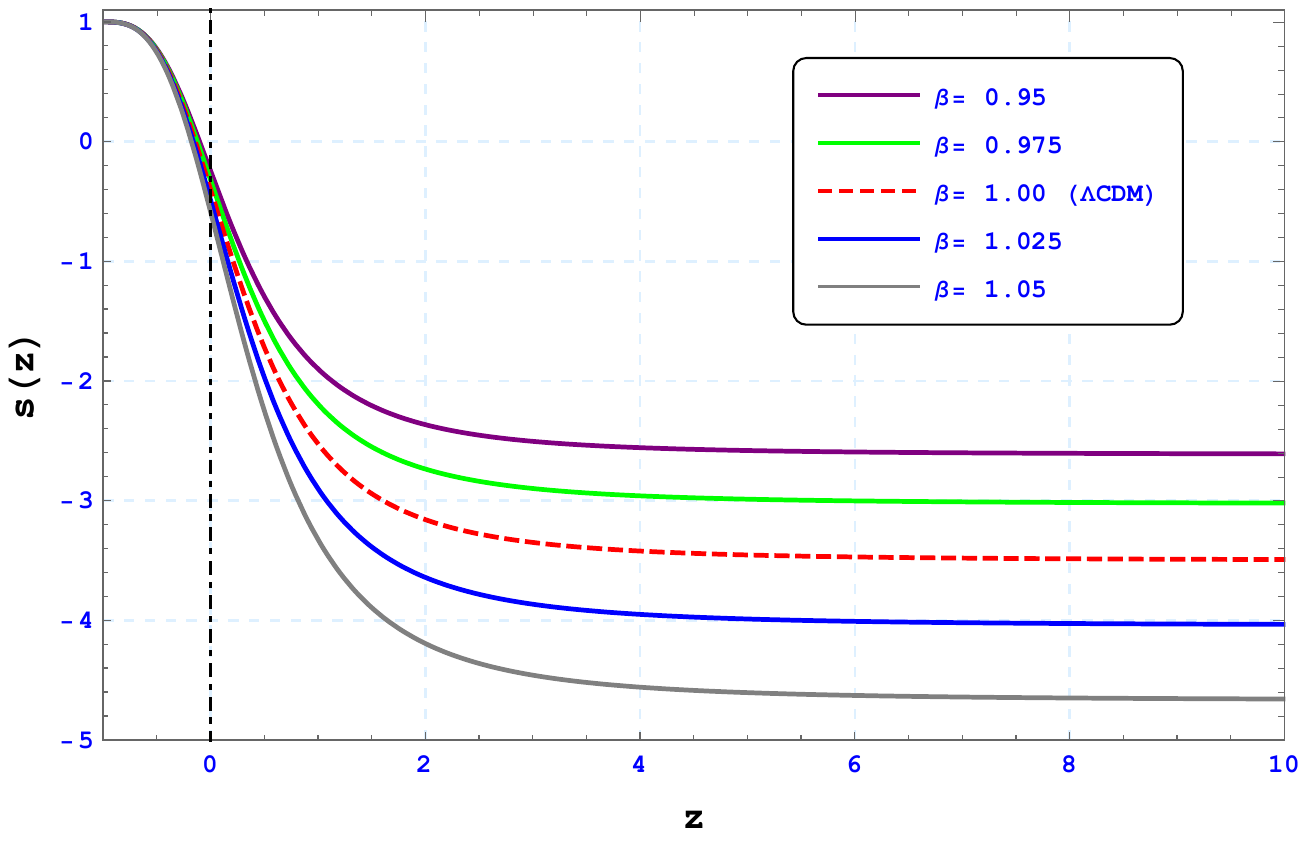}
    \caption{\small(Color online) Plot for snap parameter $s(z)$ versus redshift $z$ with different values of TC model parameter $\beta$}
    \label{sTC.pdf}
\end{figure}

 Now, we explore deeper into the information carried by the Hubble, deceleration, and jerk parameters. We encounter a more sophisticated cosmographic parameter called the snap parameter. This parameter provides an even deeper understanding of the dynamics of cosmic expansion. It is important to note that, whereas the deceleration parameter signifies the shift from deceleration to acceleration in our Universe, and the jerk describes the evolution of acceleration, the snap parameter illustrates the development of the jerk parameter. Therefore, the snap parameter serves as an additional important tool for investigating deviations from the standard $\Lambda$CDM model. The snap parameter \(s(z)\) in the TC-modified cosmology can be defined as
\begin{align}\label{snap_parameter}
s(z)
&\overset{\text{def}}{=} \frac{\ddddot a}{aH^4}
= -j(z)\,[3q(z)+2] - (1+z)\frac{dj(z)}{dz}
\nonumber \\[6pt]
&= -\frac{1}{
\left[\Omega_{m,0}(1+z)^3+\Omega_{\Lambda,0}\right]^3
}
 \nonumber \\[4pt]
&\qquad\quad \times
\Bigg[
\frac{2(2\beta-1)(\beta+1)(2\beta+5)}{(4-2\beta)^3}\,
\Omega_{m,0}^3(1+z)^9
\nonumber\\[4pt]
&\qquad\quad
- \frac{6(2\beta^2 -17\beta+11)}{(4-2\beta)^2}\,
\Omega_{m,0}^2\Omega_{\Lambda,0}(1+z)^6
\nonumber\\[4pt]
&\qquad\quad
+ \frac{3(2\beta-1)}{(4-2\beta)}\,
\Omega_{m,0}\Omega_{\Lambda,0}^2(1+z)^3
- \Omega_{\Lambda,0}^3
\Bigg]\,.
\end{align}
We have illustrated the evolution of the snap parameter with redshift for various values of the model parameter \(\beta\) in Fig. \ref{sTC.pdf}. We observe that the snap parameter decreases with an increase in \(\beta\). For the flat \(\Lambda\)CDM model, the snap parameter starts at \(s=-3.5\) during the matter-dominated phase, reaches \(s=-2\) at the transition time from deceleration to acceleration, and approaches the de-Sitter asymptotic value of \(s=1\), which highlights the distinction between the \(\Lambda\)CDM and the TC gravity models, except in the far future (\(z \to -1\)).
\par
\begin{figure}[htbp!]
    \centering
    \includegraphics[width=8cm,height=6cm]{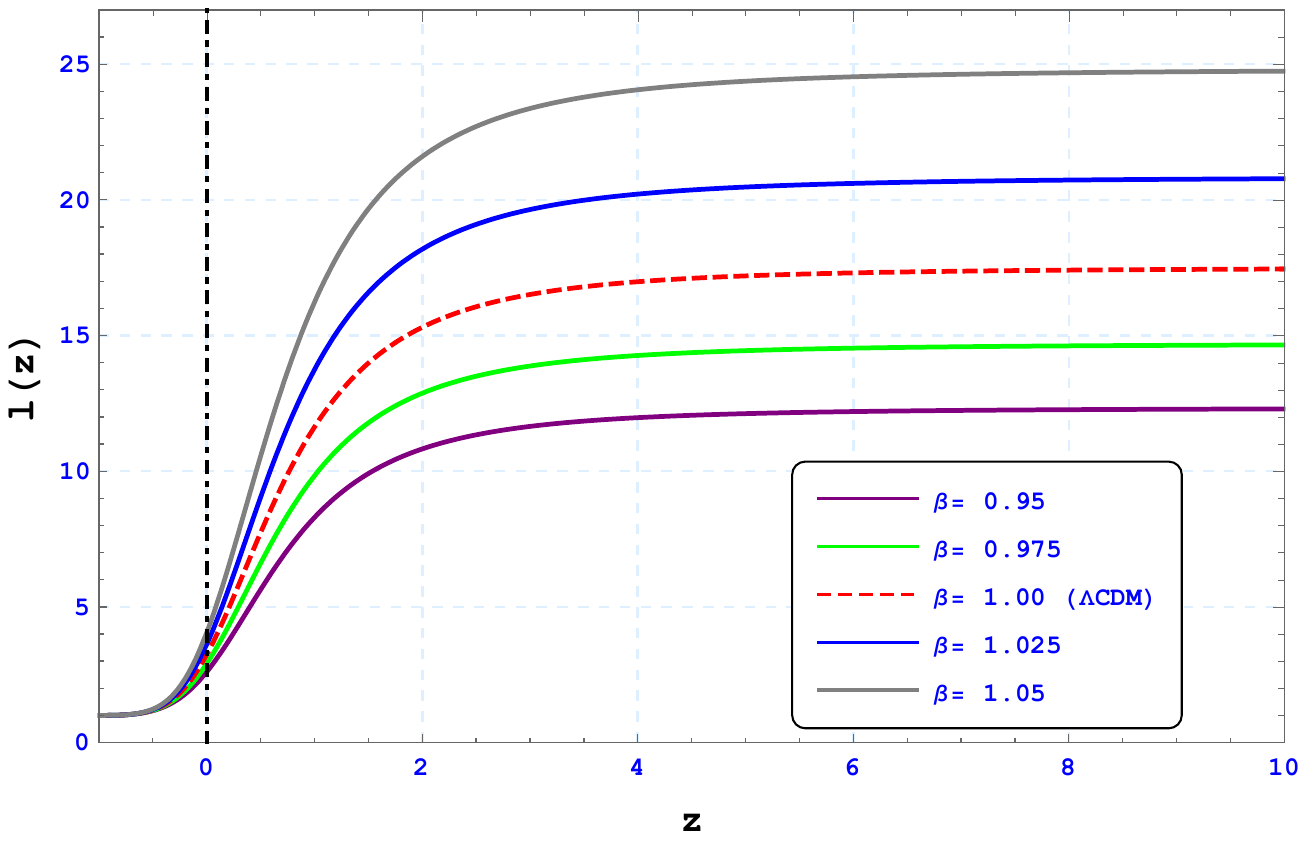}
    \caption{\small(Color online) Plot for lerk parameter $l(z)$ versus redshift $z$ with different values of TC model parameter $\beta$}
    \label{lTC.pdf}
\end{figure}

 Similarly, we present the next higher-order cosmographic metric, the lerk parameter, to quantify the rate of change of the snap parameter, which offers a delicate measure to detect subtle deviations from the $\Lambda$CDM profile. The lerk parameter $\ell (z)$ in the TC-modified cosmology is defined as
\begin{align}\label{lerk_parameter}
\ell(z)
&\overset{\text{def}}{=} \frac{\overset{\cdot\cdot\cdot\cdot\cdot}{a}}{aH^5}
= -s(z)\,[4q(z)+3] - (1+z)\frac{ds(z)}{dz}
\nonumber \\[6pt]
&=
\frac{1}{
\big[\Omega_{m,0}(1+z)^3 + \Omega_{\Lambda,0}\big]^4
}\nonumber \\[4pt]
&\qquad\quad \times
\Bigg[
\frac{4(2\beta-1)(\beta+1)(2\beta+5)(\beta+4)}{(4-2\beta)^4}\,
\Omega_{m,0}^4(1+z)^{12}
\nonumber\\[4pt]
&\qquad\quad
- \frac{8(2\beta-1)\big(2\beta^2 -65\beta +23\big)}{(4-2\beta)^3}\,
\Omega_{\Lambda,0}\Omega_{m,0}^3(1+z)^9
\nonumber\\[4pt]
&\qquad\quad
+ \frac{6\big(4\beta^2+62\beta-47\big)}{(4-2\beta)^2}\,
\Omega_{\Lambda,0}^2\Omega_{m,0}^2(1+z)^6
\nonumber\\[4pt]
&\qquad\quad
+ \frac{2(11-4\beta)}{(4-2\beta)}\,
\Omega_{\Lambda,0}^3\Omega_{m,0}(1+z)^3
+ \Omega_{\Lambda,0}^4
\Bigg]\,.
\end{align}
We have illustrated the evolution of the lerk parameter with redshift for different values of the model parameter $\beta$ in Fig. \ref{lTC.pdf}. We note that the higher the value of the TC parameter $\beta$, the higher the lerk parameter across all redshifts. The modified cosmological model resembles the standard $\Lambda$CDM scenario at $z \to -1$.
\par
\begin{figure}[htbp!]
    \centering
    \includegraphics[width=8cm,height=6cm]{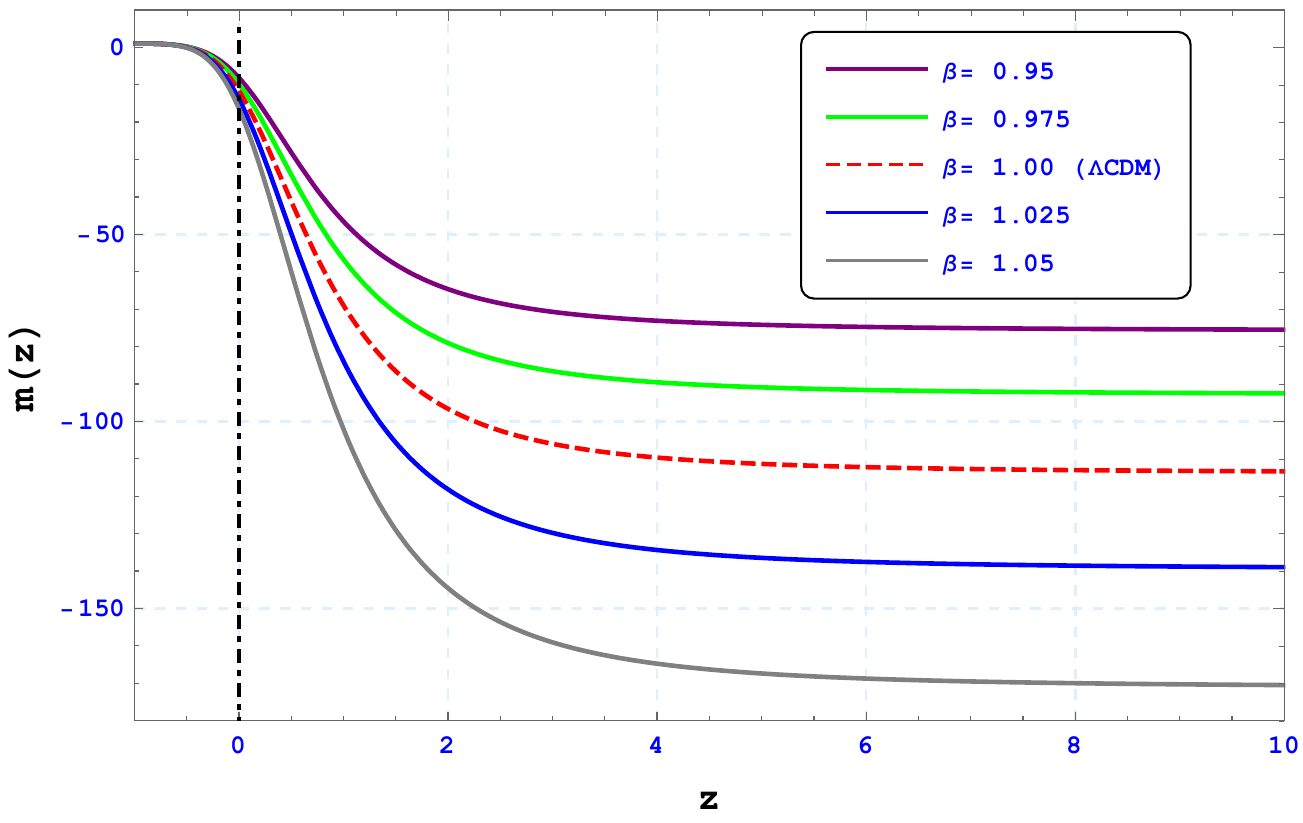}
    \caption{\small(Color online) Plot for m--parameter $l(z)$ versus redshift $z$ with different values of TC model parameter $\beta$}
    \label{mTC.pdf}
\end{figure}

 This chain of derivatives can further be extended to obtain the succeeding higher-order kinematical parameters. However, we limit ourselves up to the sixth derivative of the scale factor, termed the $m$-parameter. This documents the evolution of the lerk parameter and acts as a sensitive probe to identify deviations from the standard cosmological scenario. The $m$-parameter in the TC cosmology can be written as
\begin{widetext}
\begin{align}\label{m_parameter}
m(z)
&\overset{\text{def}}{=} \frac{\overset{\cdot\cdot\cdot\cdot\cdot\cdot}{a}}{aH^6}
= -\ell(z)\,[5q(z)+4] - (1+z)\frac{d\ell(z)}{dz}
\nonumber \\[6pt]
&=
\frac{1}{
\big[\Omega_{m,0}(1+z)^3 + \Omega_{\Lambda,0}\big]^5
}\nonumber \\[4pt]
&\qquad\quad \times
\Bigg[
-\frac{4(2\beta-1)(\beta+1)(2\beta+5)(\beta+4)(2\beta+11)}{(4-2\beta)^5}\,
\Omega_{m,0}^5(1+z)^{15}
\nonumber\\[4pt]
&\qquad\quad
+ \frac{4(2\beta-1)(10\beta^3 -1209\beta^2 +135\beta -266)}{(4-2\beta)^4}\,
\Omega_{m,0}^4\Omega_{\Lambda,0}(1+z)^{12}
\nonumber\\[4pt]
&\qquad\quad
- \frac{2(40\beta^3 +6672\beta^2 -7944\beta+2569)}{(4-2\beta)^3}\,
\Omega_{m,0}^3\Omega_{\Lambda,0}^2(1+z)^9
\nonumber\\[4pt]
&\qquad\quad
+ \frac{20(2\beta^2 -125\beta +98)}{(4-2\beta)^2}\,
\Omega_{m,0}^2\Omega_{\Lambda,0}^3(1+z)^6
\nonumber\\[4pt]
&\qquad\quad
- \frac{(7+10\beta)}{(4-2\beta)}\,
\Omega_{m,0}\Omega_{\Lambda,0}^4(1+z)^3
+ \Omega_{\Lambda,0}^5
\Bigg]\,.
\end{align}
\end{widetext}
The redshift evolution of the $m$ parameter for different values of the TC model parameter $\beta$ is shown in Fig. \ref{mTC.pdf}. We observe that the $m$ parameter decreases with the increase of the model parameter $\beta$, and it is always negative throughout all the redshifts. Again, the modified model matches the $\Lambda$CDM profile at $z \to -1$.
\begin{figure}[htbp!]
    \centering
    \includegraphics[width=8cm,height=6cm]{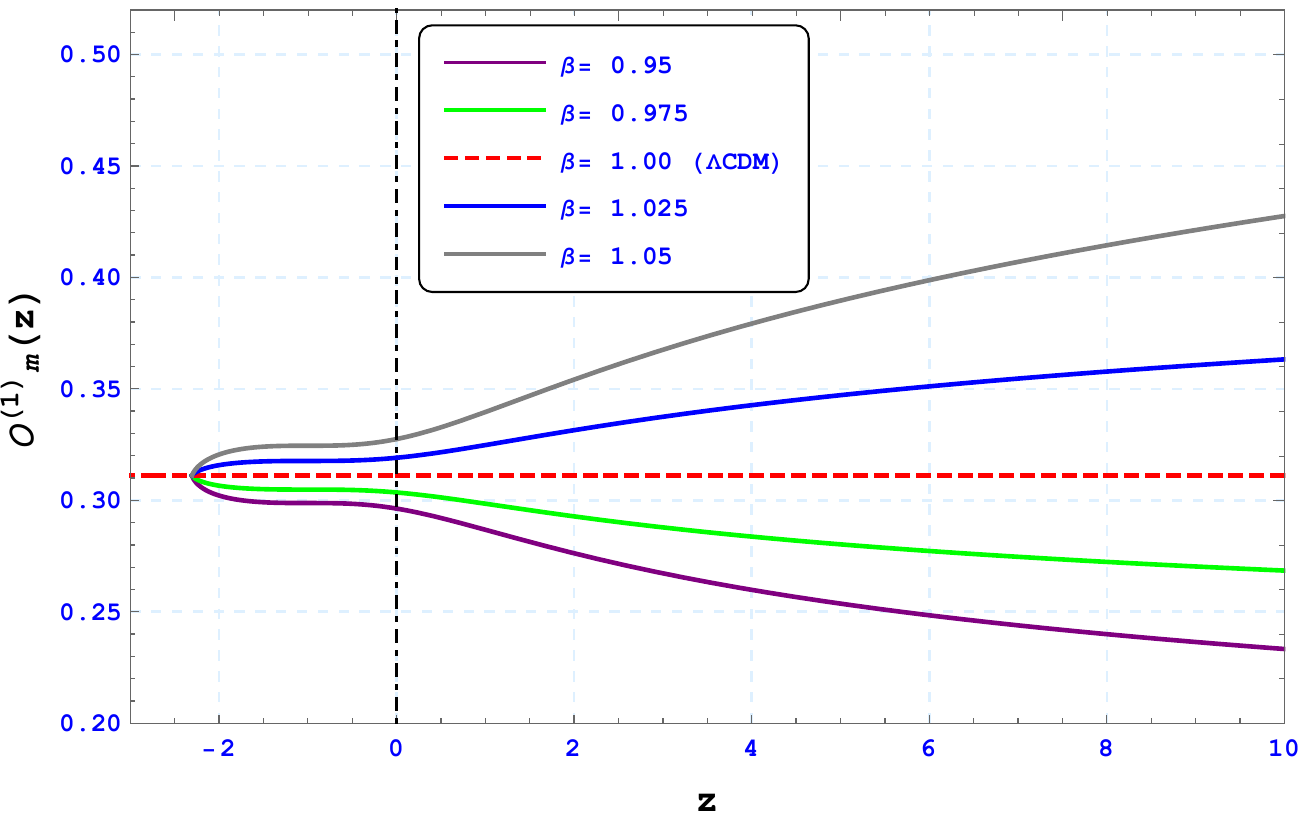}
    \caption{\small(Color online) Plot for $\mathcal{O}^{(1)}_m(z)$ parameter versus redshift $z$ with different values of TC model parameter $\beta$}
    \label{O1mTC.pdf}
\end{figure}
\begin{figure}[htbp!]
    \centering
    \includegraphics[width=8cm,height=6cm]{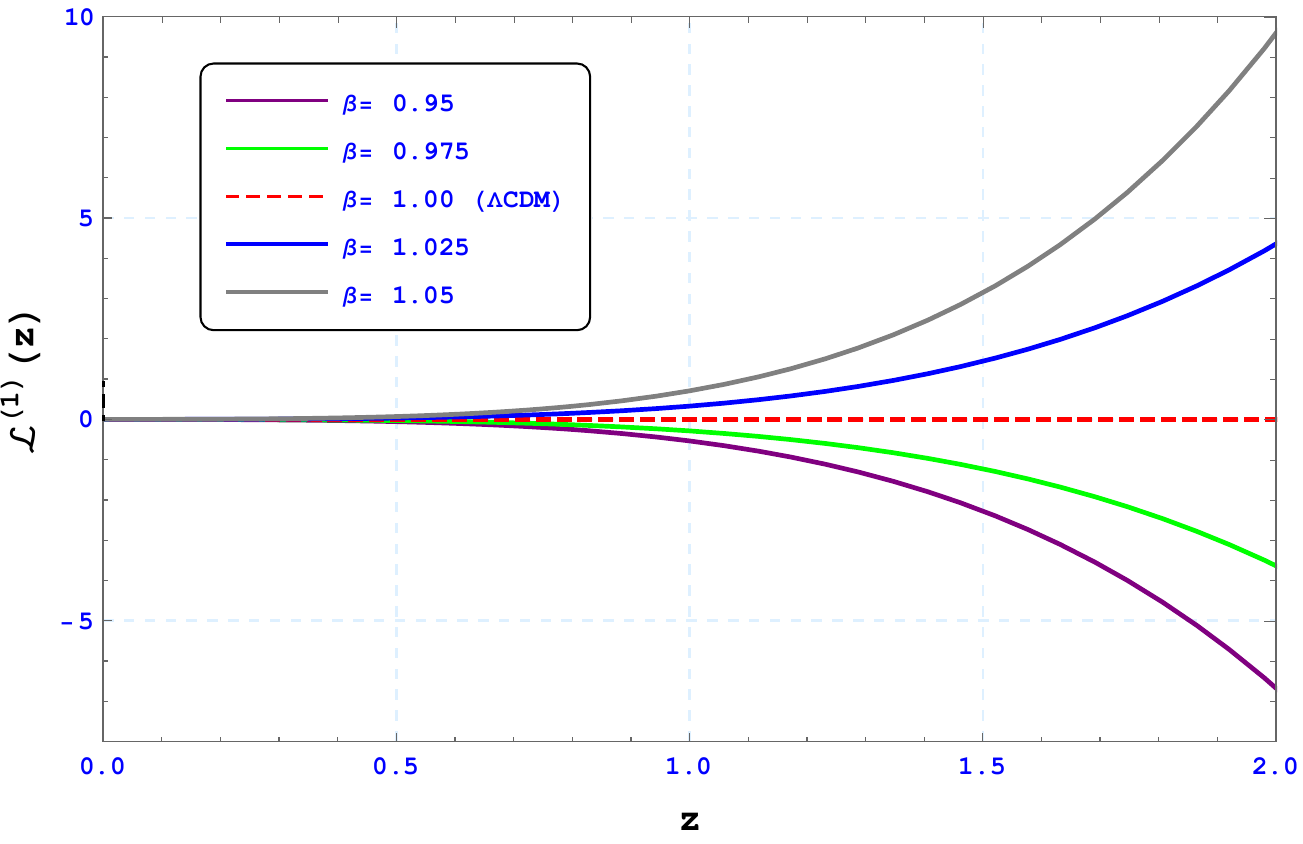}
    \caption{\small(Color online) Plot for $\mathcal{L}^{(1)}(z)$ parameter versus redshift $z$ with different values of TC model parameter $\beta$}
    \label{L1TC.pdf}
\end{figure}
\begin{figure}[htbp!]
    \centering
    \includegraphics[width=8cm,height=6cm]{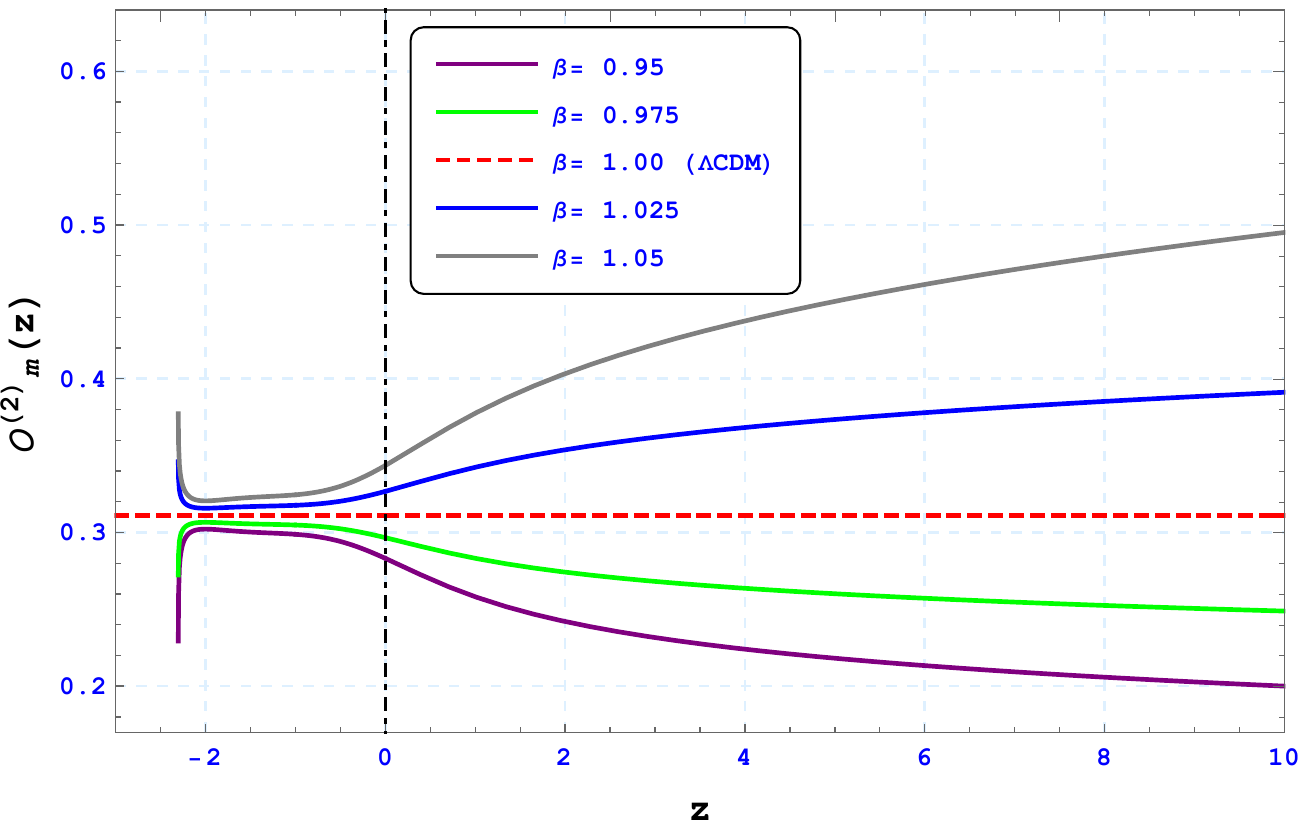}
    \caption{\small(Color online) Plot for $\mathcal{O}^{(2)}_m(z)$ parameter versus redshift $z$ with different values of TC model parameter $\beta$}
    \label{O2mTC.pdf}
\end{figure}
\begin{figure}[htbp!]
    \centering
    \includegraphics[width=8cm,height=6cm]{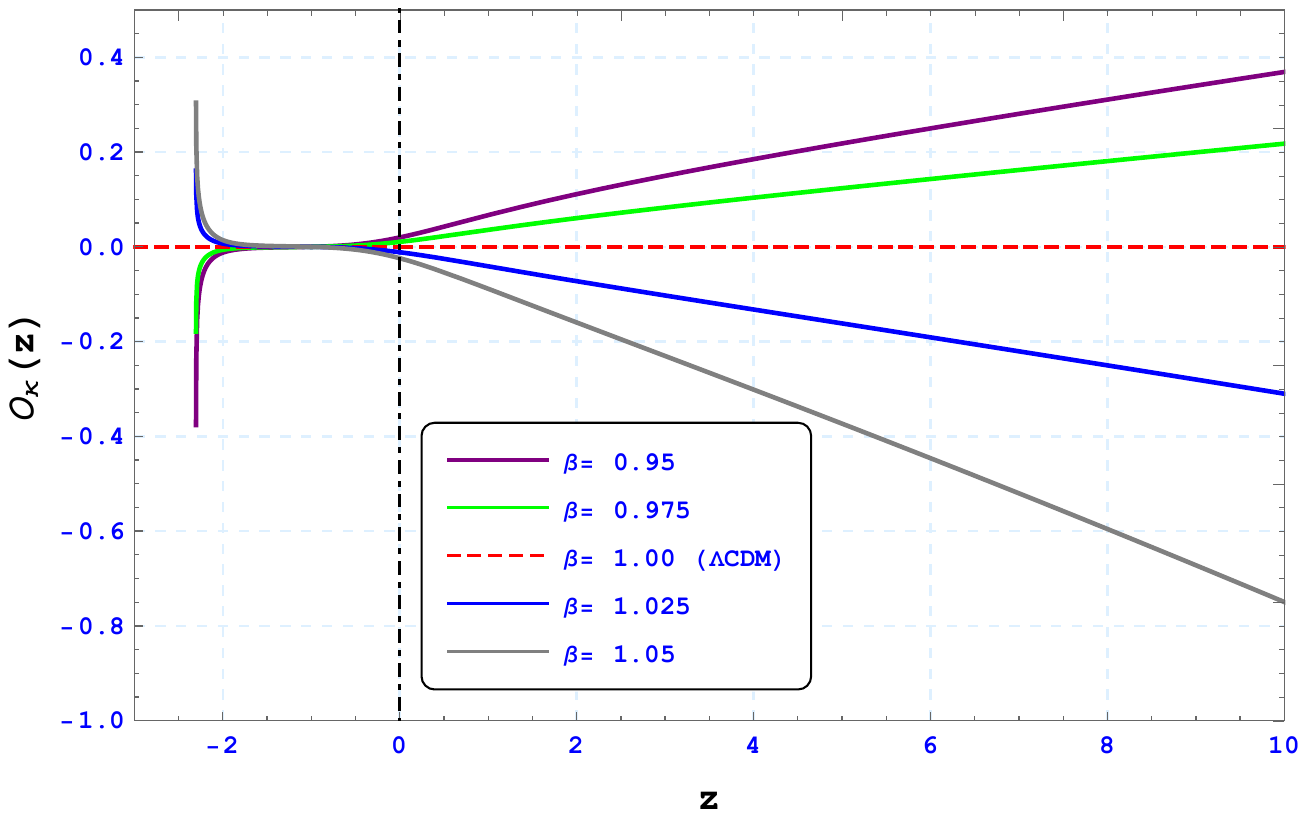}
    \caption{\small(Color online) Plot for $\mathcal{O}_\kappa(z)$ parameter versus redshift $z$ with different values of TC model parameter $\beta$}
    \label{OkTC.pdf}
\end{figure}
\begin{figure}[htbp!]
    \centering
    \includegraphics[width=8cm,height=6cm]{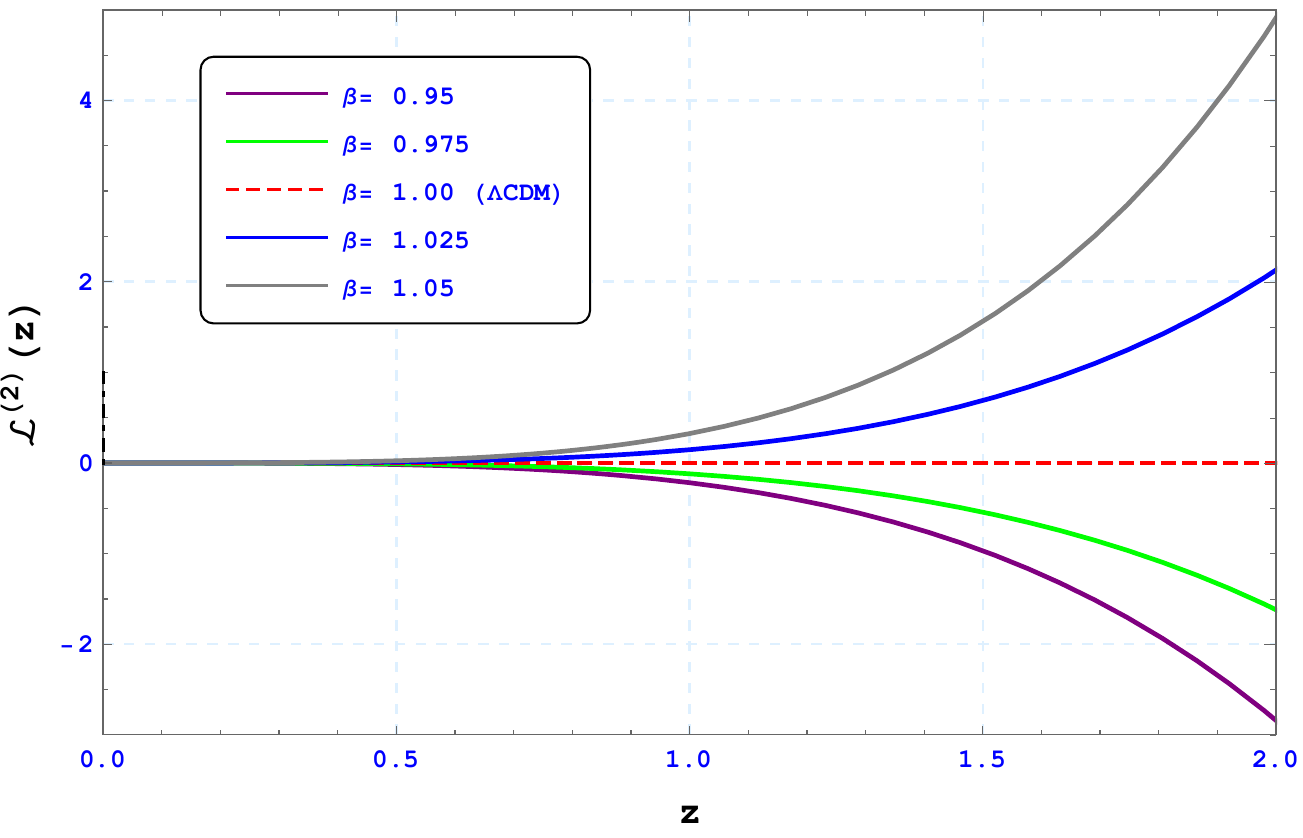}
    \caption{\small(Color online) Plot for $\mathcal{L}^{(2)}(z)$ parameter versus redshift $z$ with different values of TC model parameter $\beta$}
    \label{L2TC.pdf}
\end{figure}
\par In addition to cosmography, a novel and recognized geometric diagnostic is discussed in Refs. \cite{sahni2008two,seikel2012using} to differentiate different cosmological models in relation to flat and non-flat $\Lambda$CDM frameworks. In the TC-modified cosmological framework, the first diagnostic parameter $\mathcal{O}^{(1)}_m(z)$ can be expressed as
\begin{align}\label{Om(z)_param}
       \mathcal{O}^{(1)}_m(z) &\overset{\text{def}}{=} \frac{E^2(z)-1}{(1+z)^3-1} \nonumber \\[6pt]
       &= \frac{[\Omega_{m,0}(1+z)^{3}+\Omega_{\Lambda,0}]^{\frac{1}{2-\beta}}-1}{(1+z)^3-1} \,.  
\end{align}
When $\mathcal{O}^{(1)}_m(z)=\Omega_{m,0}$, it implies either the spatially flat $\Lambda$CDM (the concordance model), or DE $\equiv\Lambda$. From Eq. (\ref{Om(z)_param}), we obtain that $\mathcal{O}^{(1)}_m(z)\big\rfloor_{\beta = 1} = \Omega_{m,0}$ holds for any redshift in the TC model, thereby reaffirming that $\beta=1$ reverts to the flat $\Lambda$CDM paradigm. The redshift dependence of $\mathcal{O}^{(1)}_m(z)$ is illustrated in Fig. \ref{O1mTC.pdf} for different values of the TC model parameter $\beta$. We note that $\mathcal{O}^{(1)}_m(z)$ is not constant for $\beta \ne 1$ cases. This suggests a modified DE or an alternative gravity framework and excludes the standard concordance model for cases when $\beta \ne 1$. The alteration in cosmology through nonextensive entropy makes an impact on the constant nature of DE in our consideration. Whenever $\mathcal{O}^{(1)}_m(z) < \Omega_{m,0}$ (for $\beta <1$), this model demonstrates phantom-like behaviour and $\mathcal{O}^{(1)}_m(z) > \Omega_{m,0}$ (for $\beta >1$) indicates quintessence-like behaviour. A more efficient formulation to measure the deviation from null instead of a constant value is to form a parameter $\mathcal{L}^{(1)}(z)$ from the first derivative with respect to redshift, $\mathcal{O}^{(1)\prime}_m(z)$ \cite{zunckel2008consistency,seikel2012using}. In the TC-modified cosmological scenario, one obtains
\begin{align}
       \mathcal{L}^{(1)}(z) &\overset{\text{def}} {=} 3(1+z)^{2}(1-E(z)^{2}) + 2z(3+3z+z^{2}) E(z) E'(z)\nonumber \\[6pt]
       &= 3(1+z)^2 \left[1- \left\{  \Omega_{m0} (z+1)^3 + \Omega_{\Lambda 0} \right\}^{\frac{1}{2-\beta}} \right. \nonumber \\[4pt]
        & \quad \left. +\frac{2z(z^2+3z+3) \Omega_{m0}\left\{\Omega_{m0} (z+1)^3+\Omega_{\Lambda 0} \right\}^{-\frac{1-\beta}{2-\beta}}}{(4-2\beta)} \right]\,.
\end{align}
The null test informs that $\mathcal{L}^{(1)}(z) \ne 0$ counterfeit the concordance model. In Fig. \ref{L1TC.pdf}, we display $\mathcal{L}^{(1)}(z)$ versus redshift for different $\beta$. It clearly illustrates a non-constant deviation from zero for $\beta \ne 1$ across all redshifts, which supports the claim made by the $\mathcal{O}^{(1)}_m$ diagnostics. As a result, the concordance model is dismissed for $\beta \ne 1$ based on both the $\mathcal{O}^{(1)}_m$ and $\mathcal{L}^{(1)}$ diagnosis. Nevertheless, there is still a possibility that a non-flat $\Lambda$CDM model adequately describes the Universe \cite{seikel2012using}. In order to differentiate from the non-flat $\Lambda$CDM model, one can calculate two  additional diagnostics parameters $\mathcal{O}_{m}^{(2)}(z)$ and $\mathcal{O}_{\kappa}(z)$ for the TC-modified cosmological scenario as
\begin{align}
       \mathcal{O}_{m}^{(2)}(z) &\overset{\text{def}}{=} \frac{2 \left[ (1+z) (1-E(z)^2) + z(2+z) E(z) E'(z) \right]}{z^2 (1+z) (3+z)} \nonumber \\[6pt]
       &= \frac{2}{z^2 (3+z)} \left[ 1 - \left\{\Omega_{m0} (1+z)^3 +\Omega_{\Lambda 0}\right\}^{\frac{1}{2-\beta}} \right. \nonumber \nonumber \\[4pt]
       & \quad \left. + \frac{3z(1+z)(2+z) \Omega_{m0}\left\{\Omega_{m0} (1+z)^3+\Omega_{\Lambda 0}\right\}^{-\frac{1-\beta}{2-\beta}}}{(4-2\beta)} \right]\,,
\end{align}
and
\begin{align}
        \mathcal{O}_{\kappa}(z) &\overset{\text{def}}{=} \frac{3(1+z)^2 [E(z)^2-1] - 2z(3+3z+z^2) E(z) E'(z)}{z^2 (1+z) (3+z)} \nonumber \\[6pt]
        &= \frac{3(1+z)}{z^2(3+z)} \left[-1+ \left\{ \Omega_{\Lambda 0} + \Omega_{m0} (1+z)^3 \right\}^{\frac{1}{2-\beta}} \right. \nonumber \\[4pt]
        & \quad \left. -\frac{2z(z^2+3z+3) \Omega_{m0}\left\{\Omega_{m0} (1+z)^3+\Omega_{\Lambda 0} \right\}^{-\frac{1-\beta}{2-\beta}}}{(4-2\beta)} \right]\,,
\end{align}
respectively. $\mathcal{O}_{m}^{(2)}(z)=\Omega_{m0}$ and \footnote{In principle, this applies to non-flat $\Lambda$CDM model. However, since we consider spatially flat Universe ($\kappa = 0$), $\Omega_{\kappa}=0$ always.}$\mathcal{O}_{\kappa}(z)=\Omega_{\kappa}$ indicates a non-flat $\Lambda$CDM framework. If these parameters do not align with the aforementioned standard $\Lambda$CDM values, we can effectively dismiss the non-flat $\Lambda$CDM configuration and conclude that either DE $\neq\Lambda$ or gravity is modified. We show $\mathcal{O}_{m}^{(2)}(z)$ and $\mathcal{O}_{\kappa}(z)$ against redhift for different $\beta$ in Fig. \ref{O2mTC.pdf} and Fig. \ref{OkTC.pdf}, respectively. We see non-constant deviations from their prescribed $\Lambda$CDM values for $\beta \ne 1$ cases, indicating a modified gravity scenario. Akin $\mathcal{L}^{(1)}(z)$, a more effective method for measuring the deviation from zero, rather than a constant value, is to define a parameter $\mathcal{L}^{(2)}(z)$ from the first derivative $\mathcal{O}^{(2)\prime}_m(z)$ or $\mathcal{O^\prime}_{\kappa}(z)$ with respect to redshift \cite{zunckel2008consistency,seikel2012using}. In the TC-modified cosmological scenario, one obtains 
\begin{align}
       \mathcal{L}^{(2)}(z) &\overset{\text{def}}{=} 3(1+z)^{2}(E(z)^{2}-1) - 2z(3+6z+2z^{2}) E(z) E'(z) \nonumber \\
       & \quad + z^{2}(3+z)(1+z) \left[ E'(z)^{2} + E(z) E''(z) \right] \nonumber \\[10pt]
       &= 3(1+z)^2 \Bigg[-1+ \left\{ \Omega_{m0} (1+z)^3 +\Omega_{\Lambda 0} \right\}^{\frac{1}{2-\beta}} \nonumber \\
       & \quad + \frac{z\Omega_{m0} \left\{ \Omega_{m0} (1+z)^3 +\Omega_{\Lambda 0} \right\}^{-\frac{3-2\beta}{2-\beta}}}{(4-2\beta)^2} \nonumber \\
       & \quad \times \Big\{\Omega_{m0}(1+z)^3 \Big\{ (2\beta-1)(5z^2+15z+6) \nonumber \\
       & \quad - 9(1+z)(2+z) \Big\} -2(4-2\beta)\Omega_{\Lambda 0}(z^2+3z+3) \Big\} \Bigg]\,.
\end{align}
The null test indicates that when $\mathcal{L}^{(2)}(z) \ne 0$, it falsifies the non-flat $\Lambda$CDM Universe. In Fig. \ref{L2TC.pdf}, we display redshift evolution of $\mathcal{L}^{(2)}(z)$ for different $\beta$ values. It clearly illustrates a non-constant deviation from zero for $\beta \ne 1$ at all redshifts, which agrees with the assertion regarding $\mathcal{O}^{(2)}_m$ as well as $\mathcal{O}_{\kappa}$ diagnostics. Thus, the non-flat $\Lambda$CDM model is dismissed for $\beta \ne 1$ based on $\mathcal{O}^{(2)}_m$, $\mathcal{O}_{\kappa}$ as well as $\mathcal{L}^{(2)}$ diagnostics. The TC-modified cosmology ($\beta \ne 1$) effectively qualifies all the litmus tests by disproving both the flat and non-flat $\Lambda$CDM frameworks.
\section{Growth of matter spherical overdensities in the Tsallis-Cirto modified cosmology}\label{Growth of matter spherical overdensities in TC-modified cosmology}
\par We consider a background Universe that consists of non-relativistic pressureless dust matter (including both visible matter and CDM), i.e., $p=p_m=0$, along with DE represented by the cosmological constant. The continuity Eq. (\ref{continuity_equation}) for the matter sector is given by
\begin{equation}\label{continuity_eqn_background}
    \dot{\rho}_m + 3H\rho_m = 0\,.
\end{equation}
To analyze the growth of spherical perturbations of matter (DE perturbations are not taken into account), we consider a spherically symmetric perturbed cloud of radius $a_c$, filled with homogeneous matter density characterized by $\rho^c_m$. Applying the top-hat SC formalism, this spherical region is defined with a uniform density and a top-hat profile, allowing us to write $\rho^c_m(t) = \rho_m(t) + \delta\rho_m(t)$ at any time $t$. If $\delta\rho_m(t) > 0$, this spherical region will eventually collapse due to its own gravitational pull. In contrast, if $\delta\rho_m(t) < 0$, it will expand more rapidly than the average Hubble growth rate, thereby forming an underdense region (void) \cite{farsi2022structure,farsi2023evolution,ziaie2020structure}. The assumption of a top-hat profile enhances the convenience of the SC model, as the uniformity of the perturbation is maintained during the entire collapse process. This results in a purely time-dependent evolution, rather than being space-dependent. As a result, one can avoid gradients within the perturbed region. Therefore, the top-hat SC model effectively represents the evolution of a uniform mini-Universe within a broader, uniform Universe \cite{fernandes2012spherical}. In this regard, it is important to note that the growth of overdense regions is slower in comparison to the remainder of the Universe during the matter-dominated epoch of the Universe. This implies that if their density reaches a sufficiently high value, they eventually collapse into galaxy clusters and other gravitationally bound structures \cite{Ryden:1970vsj}. Similar to the background Eq. (\ref{continuity_eqn_background}), the continuity equation for the matter within the spherically perturbed region of radius $a_c$ and with a local expansion rate $H_c=\frac{\dot a_c}{a_c}$ can be expressed as
\begin{equation}\label{continuity_eqn_peturbation}
    \dot{\rho}^c_m + 3H_c\rho^c_m = 0\,.    
    \end{equation}
Now, to analyze the growth of overdensities, we define a dimensionless quantity, called the matter density contrast (MDC) 
    \begin{equation}\label{density_contrast}
    \delta_m = \frac{\rho^c_m - \rho_m}{\rho_m}
    = \frac{\delta \rho_m}{\rho_m} \, ,
    \end{equation}
which measures the difference between densities of the local and background matter fluids. The first- and second-order time derivatives of MDC are given by
    \begin{equation}\label{first derivative_density contrast}
    \dot{\delta}_m = 3(1+\delta_m)(H - H_c) \,,
    \end{equation}
    \begin{equation}\label{second derivative_density contrast}
    \ddot{\delta}_m
     = 3(\dot{H} - \dot{H_c})(1+\delta_m)
    + \frac{\dot{\delta}_m^{\,2}}{1+\delta_m} \,.
\end{equation}
In obtaining the above Eqs. (\ref{first derivative_density contrast}) and (\ref{second derivative_density contrast}), we have exploited Eqs. (\ref{continuity_eqn_background}) and (\ref{continuity_eqn_peturbation}). Merging Eqs. (\ref{contituity_eqn_seperate}), with Eq. (\ref{mod_Friedmann_eqn}) and its derivative, and using $\dot{H}=\frac{\ddot{a}}{a} - H^2$, we get for the background Universe
\begin{equation}\label{background_evolution}
    \frac{\ddot{a}}{a} = \frac{(1-2\beta)\, \Gamma_{\beta}^\frac{1}{2-\beta}}{4-2\beta}(\rho_m + \rho_\Lambda)^\frac{1}{2-\beta} + \frac{3\, \Gamma_{\beta}^\frac{1}{2-\beta}}{4-2\beta}(\rho_m + \rho_\Lambda)^{-\frac{1-\beta}{2-\beta}}\rho_\Lambda \, .
\end{equation}
According to SC formalism, a homogeneous spherically perturbed cloud of radius $a_c$ can itself be defined by identical equations that describe the evolution of the background Universe characterized by scale factor $a$ \cite{peebles2020principles}. Consequently, for the spherically uniformly perturbed region with a radius $a_c$, it follows a similar form to Eq. (\ref{background_evolution}), particularly
\begin{equation}\label{perturbed_evolution}
    \frac{\ddot{a}_c}{a_c} = \frac{(1-2\beta)\, \Gamma_{\beta}^\frac{1}{2-\beta}}{4-2\beta}(\rho^c_m + \rho_\Lambda)^\frac{1}{2-\beta} + \frac{3\, \Gamma_{\beta}^\frac{1}{2-\beta}}{4-2\beta}(\rho^c_m + \rho_\Lambda)^{-\frac{1-\beta}{2-\beta}}\rho_\Lambda \, .
\end{equation}
Generally, one expects $\beta$ and $\rho_\Lambda$ (or $\Lambda$) to be distinct for background and perturbed regions. However, for convenience, we consider that they are the same, i.e., $\beta_c = \beta$, and $\rho^c_\Lambda = \rho_\Lambda$ (or $\Lambda_c = \Lambda$). The term $(\dot{H} - \dot{H_c})$ in Eq. (\ref{second derivative_density contrast}) can be expressed using Eqs. (\ref{density_contrast}), (\ref{background_evolution}), and (\ref{perturbed_evolution}) as
\begin{align}\label{Hdot-hdot}
    \dot{H} - \dot{H_c} &= -H^2 + H_c^2 + \frac{2(2\beta-1)\, \Gamma_{\beta}^\frac{1}{2-\beta}}{(4-2\beta)^2}\frac{\rho_m}{(\rho_m + \rho_\Lambda)^{\frac{1-\beta}{2-\beta}}}\delta_m  \nonumber \\[4pt]
    &\qquad\quad
    + \frac{6(1-\beta)\, \Gamma_{\beta}^\frac{1}{2-\beta}}{(4-2\beta)^2}\frac{\rho_m\,\rho_\Lambda}{(\rho_m + \rho_\Lambda)^{\frac{3-2\beta}{2-\beta}}}\delta_m \, .
\end{align}
Since, we are operating under the linear approximation where $\delta_m <1$, we ignore $\mathcal{O}(\delta^2_m)$ and $\mathcal{O}(\dot{\delta}^2_m)$ when evaluating Eq. (\ref{Hdot-hdot}). With the same rationale, the final term in Eq. (\ref{second derivative_density contrast}) can also be disregarded. By combining Eq. (\ref{Hdot-hdot}) with Eq. (\ref{second derivative_density contrast}), and using Eq. (\ref{first derivative_density contrast}), we derive the linear differential equation that describes the temporal evolution of MDC
\begin{align}\label{temoral evolution eqn}
\ddot{\delta}_m
+ 2H \dot{\delta}_m
&- \frac{6(2\beta-1)\, \Gamma_{\beta}^{\frac{1}{2-\beta}}}{(4-2\beta)^2}
\frac{\rho_m}{(\rho_m + \rho_\Lambda)^{\frac{1-\beta}{2-\beta}}}\,\delta_m
\nonumber \\[4pt]
&- \frac{18(1-\beta)\, \Gamma_{\beta}^{\frac{1}{2-\beta}}}{(4-2\beta)^2}
\frac{\rho_m\,\rho_\Lambda}
{(\rho_m + \rho_\Lambda)^{\frac{3-2\beta}{2-\beta}}}\,\delta_m
= 0 \, .
\end{align}
In order to study the evolution of MDC regarding the redshift parameter, we first replace the time derivatives with the scale factor. It can be readily shown that
\begin{subequations}\label{time to scale factor derivatives}
    \begin{align}
       \dot{\delta}_m &= aH\delta^\prime_m \,,\\
       \ddot{\delta}_m &= aH^2\delta^{\prime\prime}_m + a\left(\frac{\ddot{a}}{a}\right)\delta^\prime_m\, . 
    \end{align}
\end{subequations}
Here, primes symbolize the order of derivatives with respect to the scale factor $a$. Thus, making use of Eqs. (\ref{Hubble}), (\ref{background_evolution}) and (\ref{time to scale factor derivatives}) in Eq. (\ref{temoral evolution eqn}), we get
\begin{align}\label{scale factor evolution equation_one}
\delta_m^{\prime\prime}
&+ \frac{3}{(4-2\beta)a}
\left\{
(3-2\beta)
+ \frac{\Gamma_{\beta}}{H^{4-2\beta}}\,\rho_\Lambda
\right\}
\delta_m^{\prime}
\nonumber \\[4pt]
&- \frac{6(2\beta-1)}{(4-2\beta)^2}
\frac{\Gamma_{\beta}}{a^2 H^{4-2\beta}}\,
\rho_m\,\delta_m
\nonumber \\[4pt]
&- \frac{18(1-\beta)}{(4-2\beta)^2}
\frac{\Gamma_{\beta}^2}{a^2 H^{2(4-2\beta)}}\,
\rho_m\,\rho_\Lambda\,\delta_m
= 0 \, .
\end{align}
Furthermore, the aforementioned Eq. (\ref{scale factor evolution equation_one}) can be simplified by including the density parameters presented in Eq. (\ref{density_parameters}), yielding
\begin{align}\label{scale factor evolution equation_two}
\delta_m^{\prime\prime}
&+ \frac{3}{(4-2\beta)a}
\left\{
(3-2\beta) + (1-\Omega_m)
\right\}
\delta_m^{\prime}
\nonumber \\[4pt]
&- \frac{6(2\beta-1)}{(4-2\beta)^2 a^2}\,
\Omega_m\,\delta_m
\nonumber \\[4pt]
&- \frac{18(1-\beta)}{(4-2\beta)^2 a^2}\,
\Omega_m\,(1-\Omega_m)\,\delta_m
= 0 \, .
\end{align}
It is noteworthy that the entropic exponent TC model parameter $\beta$ significantly influences the evolution equation. In case of a pure matter-dominated TC Universe ($\Omega_m \approx 1$ and $\Omega_\Lambda \approx 0$), the Eq. (\ref{scale factor evolution equation_two}) simplifies to
\begin{equation}\label{matter_domination_TC}
    \delta^{\prime\prime}_m + \frac{3(3-2\beta)}{(4-2\beta)a}\delta^{\prime}_m - \frac{6(2\beta-1)}{(4-2\beta)^2a^2}\delta_m = 0\, ,
\end{equation}
which possesses a solution in terms of redshift
\begin{equation}\label{sol_matter_dom_TC}
    \delta_m(z) = \mathcal{A}_{\beta} (1+z)^{-\frac{2\beta-1}{2-\beta}} + \mathcal{B}_{\beta} (1+z)^{\frac{3}{4-2\beta}}\, .
\end{equation}
Here $\mathcal{A}_{\beta}$, and $\mathcal{B}_{\beta}$ are arbitrary integration constants. It is important to mention that Eq. (\ref{scale factor evolution equation_two}) is valid at any redshift, regardless of the cosmological era. However, the study in Ref. \cite{sheykhi2022growth} is restricted to the early Universe, where the authors assume that $\rho_{\Lambda}/\rho_m <1$ when deriving \textit{Eq. (42)} of their article \cite{sheykhi2022growth}. Interestingly, in the absence of a cosmological constant ($\Lambda = 0$), \textit{Eq. (42)} of article \cite{sheykhi2022growth} aligns with Eq. (\ref{matter_domination_TC}) in this article. Moreover, one can note that, in the limit $\beta=1$, Eq. (\ref{matter_domination_TC}) becomes to
\begin{equation}\label{matter_domination_SC}
    \delta^{\prime\prime}_m + \frac{3}{2a}\delta^{\prime}_m - \frac{3}{2a^2}\delta_m = 0\, ,
\end{equation}
admitting a widely known solution
\begin{equation}\label{sol_matter_dom_GR}
    \delta_m(z) = \frac{\mathcal{A}_1}{(1+z)} + \mathcal{B}_1 (1+z)^{\frac{3}{2}}\, ,
\end{equation}
for a pure matter-dominated Universe in the standard cosmology (the GR limit) \cite{padmanabhan1993structure, weinberg2013gravitation,abramo2007structure}. This indicates that, in addition to the dynamics of TC-modified cosmology, the presence of DE influenced by this cosmology shall also play a vital role in comprehending the evolution of perturbations. As we desire to study the MDC in the context of redshift, we use a set of transformations
\begin{subequations}\label{scale factor to redshift derivatives}
    \begin{align}
       \delta^\prime_m &= -(1+z)^2\, \frac{d\delta_m}{dz} \,,\\
       \delta^{\prime\prime}_m &= (1+z)^4\,\frac{d^2\delta_m}{dz^2}+2(1+z)^3 \,\frac{d\delta_m}{dz}\,, 
    \end{align}
\end{subequations}
into Eq. (\ref{scale factor evolution equation_two}), which offers
\begin{widetext}
\begin{align}\label{evolution eqn}
(1+z)^2\,\frac{d^2 \delta_m}{dz^2}
&+ (1+z)
\left\{
\frac{(2\beta-1) - 3(1-\Omega_m)}{(4-2\beta)}
\right\}
\frac{d \delta_m}{dz}
- \frac{6}{(4-2\beta)^2}
\left\{
(2\beta-1)\,\Omega_m
+ 3(1-\beta)(1-\Omega_m)\,\Omega_m
\right\}
\delta_m
= 0 \, .
\end{align}
\end{widetext}
Taking into account the dynamical behavior of the cosmological matter density parameter outlined in Eq. (\ref{matter_density_parameter}), Eq. (\ref{evolution eqn}) offers the subsequent analytical solution
\begin{align}\label{sol_TOTAL_TC}
\delta_m(z)
&= \mathcal{C}_{\beta}\,
\left[
\Omega_{m,0}(1+z)^{3}
+ \Omega_{\Lambda,0}
\right]^{\frac{1}{4-2\beta}}
\nonumber \\[6pt]
&\quad
+ \frac{\mathcal{D}_{\beta}}{2}\,(1+z)^{2}
\left(
1 + \frac{\Omega_{m,0}(1+z)^{3}}{\Omega_{\Lambda,0}}
\right)^{\frac{3}{4-2\beta}}
\nonumber \\[6pt]
&\quad \times
\left[
\Omega_{m,0}(1+z)^{3}
+ \Omega_{\Lambda,0}
\right]^{-\frac{1}{2-\beta}}
\nonumber \\[6pt]
&\quad \times
{}_{2}\mathfrak{F}_{1}\!\left(
\frac{2}{3},\,
\frac{3}{(4-2\beta)};\,
\frac{5}{3};\,
-\frac{\Omega_{m,0}(1+z)^{3}}{\Omega_{\Lambda,0}}
\right)\, ,
\end{align}
where $\mathcal{C}_{\beta}$ and $\mathcal{D}_{\beta}$ are arbitrary integration constants, and a hypergeometric function $ {}_{2}\mathfrak{F}_{1}(a,b;c;z)$ is introduced with parameters $a,b$, and $c$, and a variable $z$. For the $\Lambda$CDM model with $\beta =1$, the above solution reads
\begin{align}\label{delta_solution_SC}
\delta_m(z)
&= \mathcal{C}_{1}\,
\sqrt{
\Omega_{m,0}(1+z)^3 + \Omega_{\Lambda,0}
}
\nonumber \\[6pt]
&\quad
+ \frac{\mathcal{D}_{1}\,(1+z)^2}{2\,\Omega_{\Lambda,0}}
\sqrt{1+
\frac{\Omega_{m,0}(1+z)^3}{\Omega_{\Lambda,0}}
}
\nonumber \\[6pt]
&\quad \times
{}_{2}\mathfrak{F}_{1}\!\left(
\frac{2}{3},\,
\frac{3}{2};\,
\frac{5}{3};\,
-\frac{\Omega_{m,0}(1+z)^3}{\Omega_{\Lambda,0}}
\right)\, ,
\end{align}
which resembles the solution found in Ref. \cite{mondal2026cosmological}. Since our analysis is confined to the linear regime and centred on the pre-collapsed evolution of perturbations, we adopt adiabatic initial conditions to determine the integration constants for the solution presented in Eq. (\ref{sol_TOTAL_TC})
\begin{equation}\label{initial_conditions}
    \delta_m(z)\big\rfloor_{z = z_i}=\delta^i_m \quad\text{and} \quad \frac{d\delta_m(z)}{dz}\Bigg\rfloor_{z = z_i} = -\left(\frac{2\beta-1}{2-\beta}\right)\frac{\delta^i_m}{1+z_i}\, .
\end{equation}
\begin{figure}[htbp!]
    \centering
    \includegraphics[width=8cm,height=6cm]{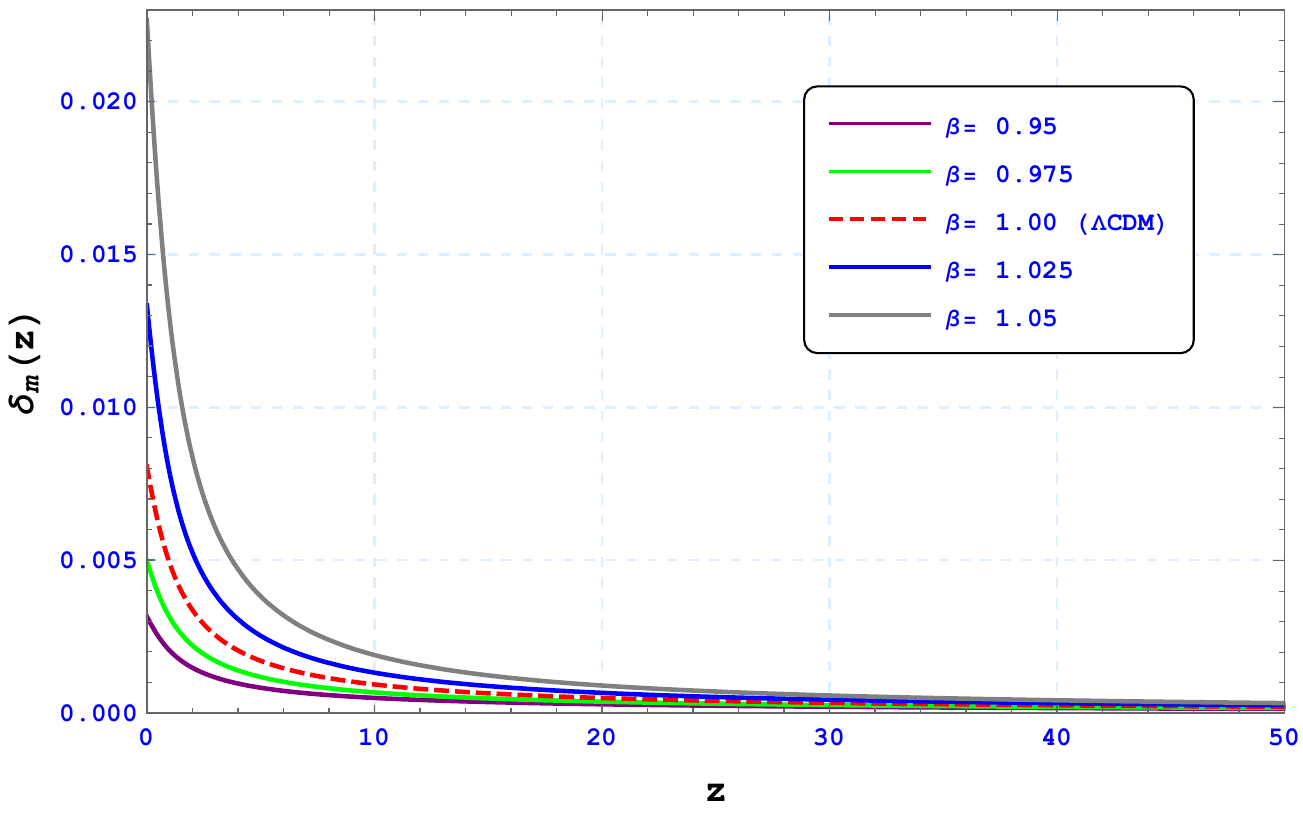}
    \caption{\small(Color online) Plot for matter density contrast $\delta_m(z)$ parameter versus redshift $z$ with different values of TC model parameter $\beta$}
    \label{delta_mTC.pdf}
\end{figure}
Specifically, we choose $z_i = 10^3$, which corresponds to a time shortly after matter-radiation equality and well before the recombination epoch, and set $\delta^i_m = 10^ {-5}$ \cite{luciano2025modified_Cos}, which matches the anticipated scale of primordial fluctuations from predictions of the inflation theory, and is supported by the CMBR anisotropy observations from the Planck mission \cite{aghanim2020planck,aghanim2021erratum}. The initial condition on the first-order derivative of the MDC arises from consideration of the negligible role of DE initially during matter domination, shortly following matter-radiation equality. To derive this expression, we focus solely on the growing mode of the solution given in Eq. (\ref{sol_matter_dom_TC}) by setting $\mathcal{B}_{\beta} = 0$, which is crucial for structure formation. In this regard, we assert that we consider only small deviations from the standard $\Lambda$CDM framework, i.e., $|\beta -1|<<1$. In Fig. \ref{delta_mTC.pdf}, we have displayed MDC $\delta_m$ against redshift $z$ for different values of the TC-entropic exponent parameter $\beta$. It is clear that the effects of TC-modified cosmology leave a significant mark on the MDC characteristics. We observe that the MDC initiates its growth from the early Universe and grows faster as the Universe expands, showing a departure from the $\Lambda$CDM profile ($\beta =1$). Additionally, we observe that the growth of perturbations intensifies with increasing $\beta$, and in particular at the lower redshifts, the impact of parameter $\beta$ becomes distinctly apparent. For $\beta<1$, the MDC exhibits a slower growth rate in contrast to standard cosmology, whereas for $\beta>1$ it demonstrates a faster growth rate.
\par
\begin{figure}[htbp!]
    \centering
    \includegraphics[width=8cm,height=6cm]{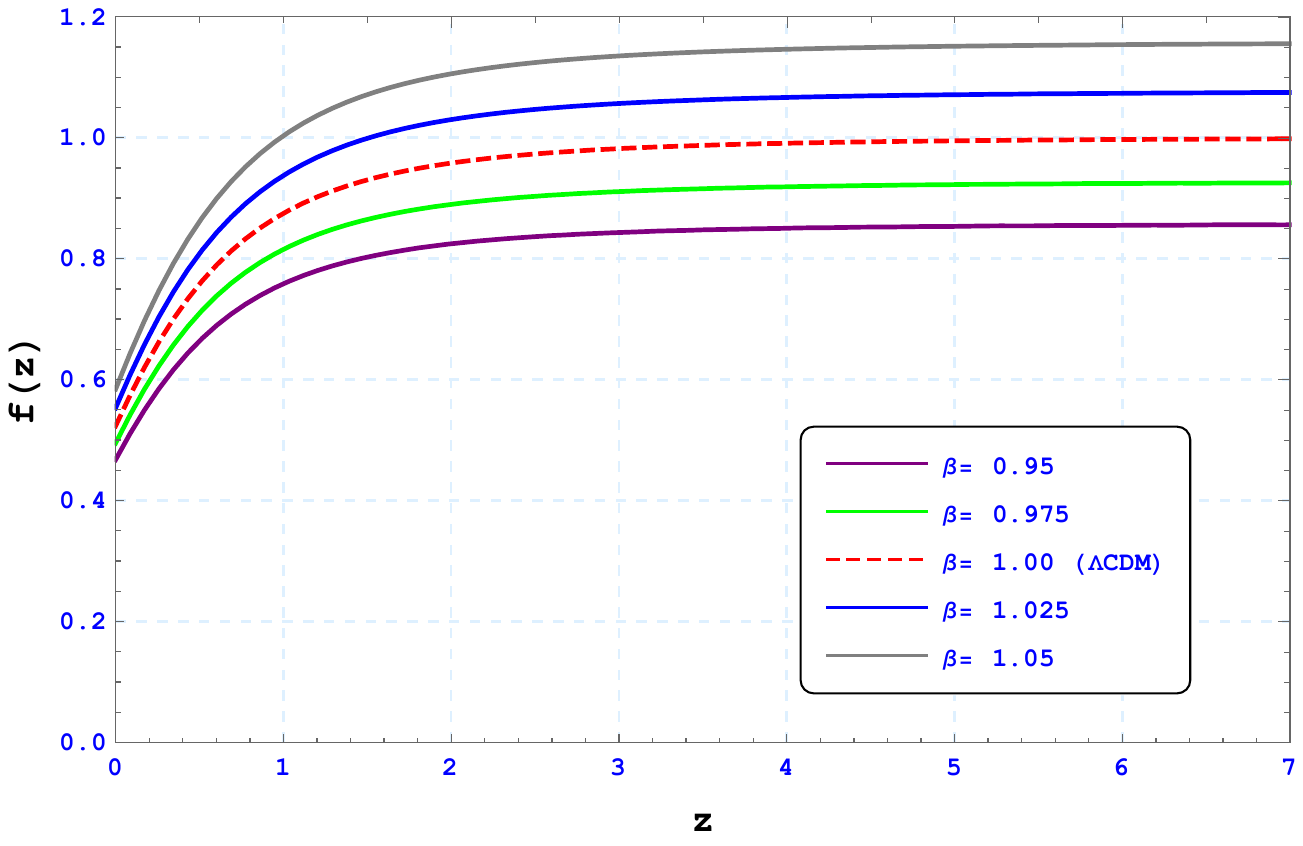}
    \caption{\small(Color online) Plot for logarithmic growth rate $f(z)$ parameter versus redshift $z$ with different values of TC model parameter $\beta$}
    \label{fTC.pdf}
\end{figure}

 In this context, we can examine the growth rate of matter overdensities using the logarithmic growth function (LGF) \cite{peebles2020principles}
\begin{equation}\label{logarithmic growth function}
    f(z) \overset{\text{def}}{=} -\frac{d\, ln \delta_m(z)}{d\, ln(1+z)} = -(1+z)\frac{1}{\delta_m(z)}\frac{d\,\delta_m(z)}{dz}\, .
\end{equation}
We have illustrated the LGF as a function of redshift for different values of $\beta$, and constructed it with the $\Lambda$CDM model in Fig. \ref{fTC.pdf}. It is evident that at high redshifts the LGF attains a saturation level (unity in standard $\Lambda$CDM cosmology), while it begins to decline at lower redshifts. The nature of $f(z)$ is significantly influenced by the TC-model parameter $\beta$, which increases with rising $\beta$. The reduction of LGF at lower redshifts is attributed to DE domination, which hampers structure formation by inhibiting growth. 
\par
\begin{figure}[htbp!]
    \centering
    \includegraphics[width=8cm,height=6cm]{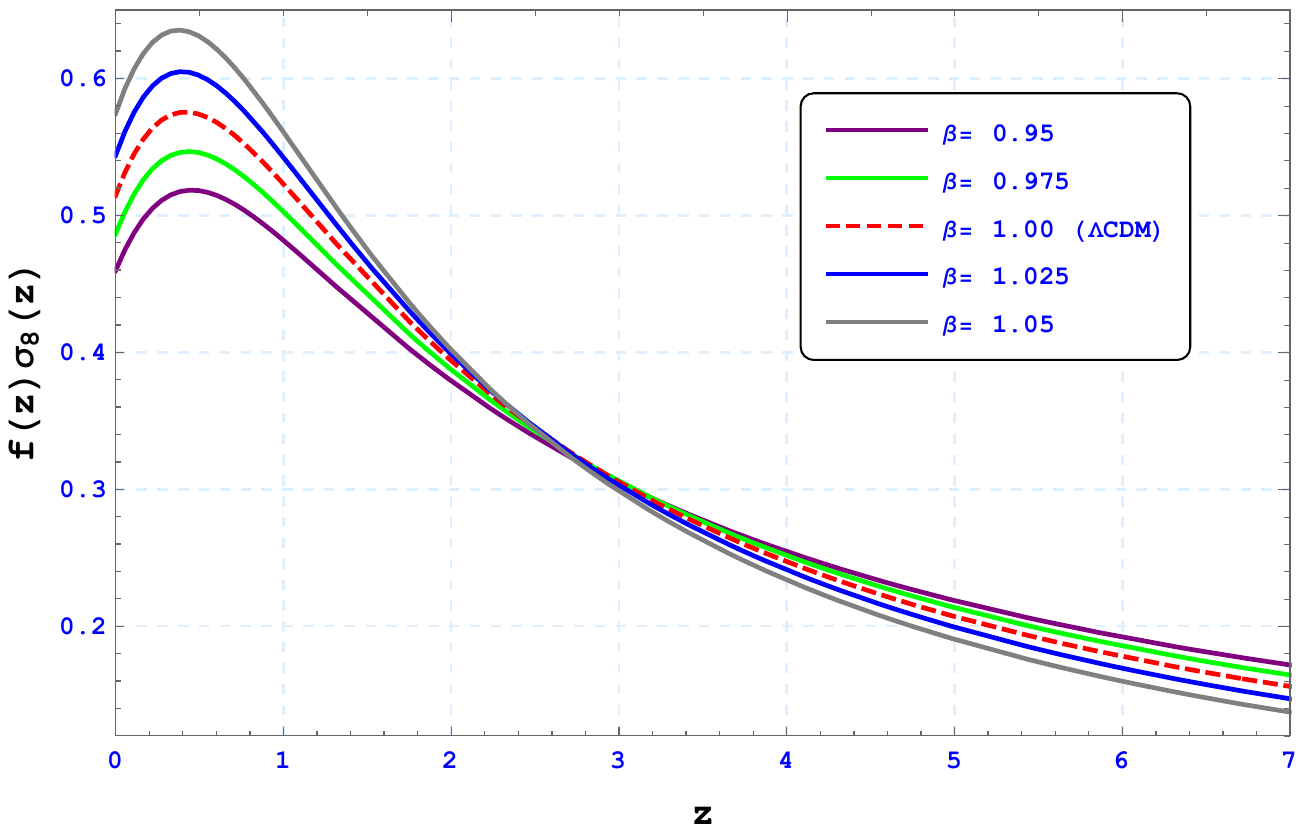}
    \caption{\small(Color online) Plot for $f(z) \sigma_8(z)$ parameter versus redshift $z$ with different values of TC model parameter $\beta$}
    \label{fsigma_8TC.pdf}
\end{figure}

 The growth rate of matter fluctuations is also assessed by examining the redshift-space distortion in the clustering pattern of galaxies. This distortion arises from the peculiar velocities linked with the inward collapse of large-scale structures, which are intimately connected with the growth rate of the MDC \cite{kaiser1987clustering}. Recent exploration on galaxy redshift serveys has contributed to establishing constraints on the LGF $f(z)$ or $f(z)\sigma_8(z)$ as a function of redshift, where $f(z)$ is derived from Eq. (\ref{logarithmic growth function}) and $\sigma_8(z)$ denotes the root mean squared amplitude of $\delta_m$ measured at a comoving scale $8 \mathrm{h}^{-1} Mpc$ \cite{tsujikawa2013testing,nesseris2017tension}, and can be defined as \cite{nesseris2008testing}
\begin{equation}
    \sigma_8(z) \overset{\text{def}}{=} \frac{\delta_m(z)}{\delta_m(z=0)}\sigma_8(z=0)\, ,
\end{equation}
assuming $\sigma_8(z=0)=0.983 \pm 0.0060$ \cite{nesseris2017tension}. The redshift evolution of $f(z)\sigma_8(z)$ for various values of the $\beta$ parameter is also displayed in Fig. \ref{fsigma_8TC.pdf}. For small redshifts ($z<1$), this TC-modified cosmological model yields a higher value of the cosmological growth rate for a larger $\beta$. However, at large redshifts, the opposite happens. Particularly, by introducing nonextensivity into the entropy within the framework of TC-modified cosmology, we observe that at low redshifts the growth rate function exceeds that of the $\Lambda$CDM model for $\beta >1$, while it falls short for $\beta <1$. Additionally, by including the model parameter $\beta$, we note that $f(z)\sigma_8(z)$ reaches its maximum value at lower redshifts for higher $\beta$ values. In essence, the large-scale structures form in a later epoch in the TC-modified cosmological model than in the standard $\Lambda$CDM model for $\beta >1$; conversely, they form earlier for $\beta <1$. This finding highlights the considerable impact of the TC-model parameter $\beta$ on the formation timing of the large-scale structures in the Universe.
\section{Halo mass function and cluster number counts in the Tsallis-Cirto modified cosmology}\label{Halo mass function and cluster number counts in TC-modified cosmology}
\par In addition to analyzing the redshift evolution of the MDC and growth functions, it would be worthwhile to investigate the number counts of the collapsed objects in the TC-modified cosmological scenario discussed in this article. It is well understood that the primary mechanism of formation of large-scale structure in the Universe involves the gravitational collapse of matter perturbations. These collapsed objects are referred to as the DM halos. The visible baryons follow the distribution of DM due to gravitational interactions. Each galaxy cluster, including our Milky Way, is thought to be embedded within one of these massive and approximately spherical clouds or halos of DM. The observation of the distribution of galaxy clusters provides insights into the distribution of DM halos in the Universe. In this section, we investigate the \textit{differential halo mass function} (DHMF) and cluster number counts of DM halos in the TC-modified cosmological framework through the Sheth-Mo-Tormen (SMT) \cite{sheth1999large,sheth2001ellipsoidal,sheth2002excursion} method, which is an enhanced version of the Press-Schechter (PS) \cite{press1974formation} formalism. 
\par Exploiting the current mean mass density $\rho_{m,0}$, root-mean-square of smoothed density fluctuation $\sigma(M,z)$ for a spherical mass $M$ with comoving radius $R$, and critical overdensity $\delta_{cr}(z)$ indicating the thresold structure collapse, we calculate the comoving number density of collapsed halos per logarithmic interval (known as DHMF) for the virial mass $M$ to $M+dM$ at a specified redshift z \cite{herrera2017calculation,gupta2022universality}
\begin{equation}\label{DHMF}
    \frac{dn}{d \ln M}(M,z) =  \frac{\rho_{m,0}}{M}  \left| \frac{d \ln \sigma(M,z)}{d \ln M} \right|\mathcal{F}(M,z)\, ,
\end{equation}
whereas the \textit{halo multiplicity function} $\mathcal{F}(M,z)$ dictates the shape of the DHMF or the mass fraction within the collapsed volume. The DHMF quantifies the comoving number density of DM halos. The abundance of DM halos is expressed as a function of their virial mass at a certain redshift, normalized to a unit volume. The value of the DHMF depends on the background Universe and the matter power spectrum. A basic theoretical model of this statistical property was proposed by Press and Schechter \cite{press1974formation}, based on fundamental principles connected to the fixed-barrier hierarchical Gaussian density fluctuation field. Subsequently, the most original formulation of the PS model was rectified through the excursion set method, also referred to as the extended PS formalism (see Ref. \cite{bond1991excursion}), which introduced the following function of $\mathcal{F}(M,z)$
\begin{equation}
    \mathcal{F}_{PS}(M,z) = \sqrt{\frac{2}{\pi}}\frac{\delta_{cr}(z)}{\sigma(M,z)}   \exp\left( -\frac{\delta_{cr}^2(z)}{2\sigma^2 (M,z)}\right)\, .
\end{equation}
\begin{figure}[htbp!]
    \centering
    \includegraphics[width=8cm,height=6cm]{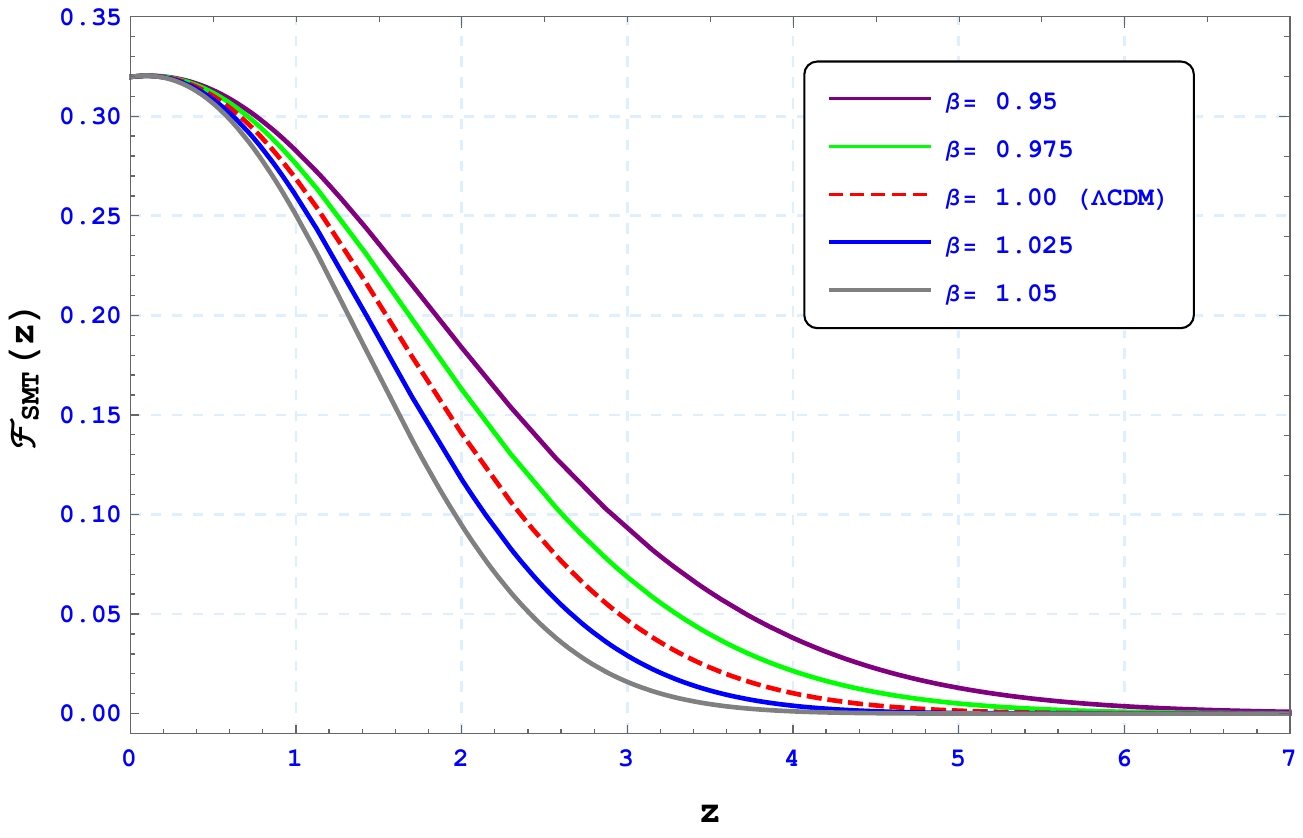}
    \caption{\small(Color online) Plot for Sheth, Mo, and Tormen halo multiplicity function $\mathcal{F}_{\mathrm{SMT}}$ versus redshift $z$ for mass $M=10^{13}\mathrm{h}^{-1}M_{\odot}$ with different values of TC model parameter $\beta$}
    \label{f_SMTTC.pdf}
\end{figure}
\begin{figure}[htbp!]
    \centering
    \includegraphics[width=8cm,height=6cm]{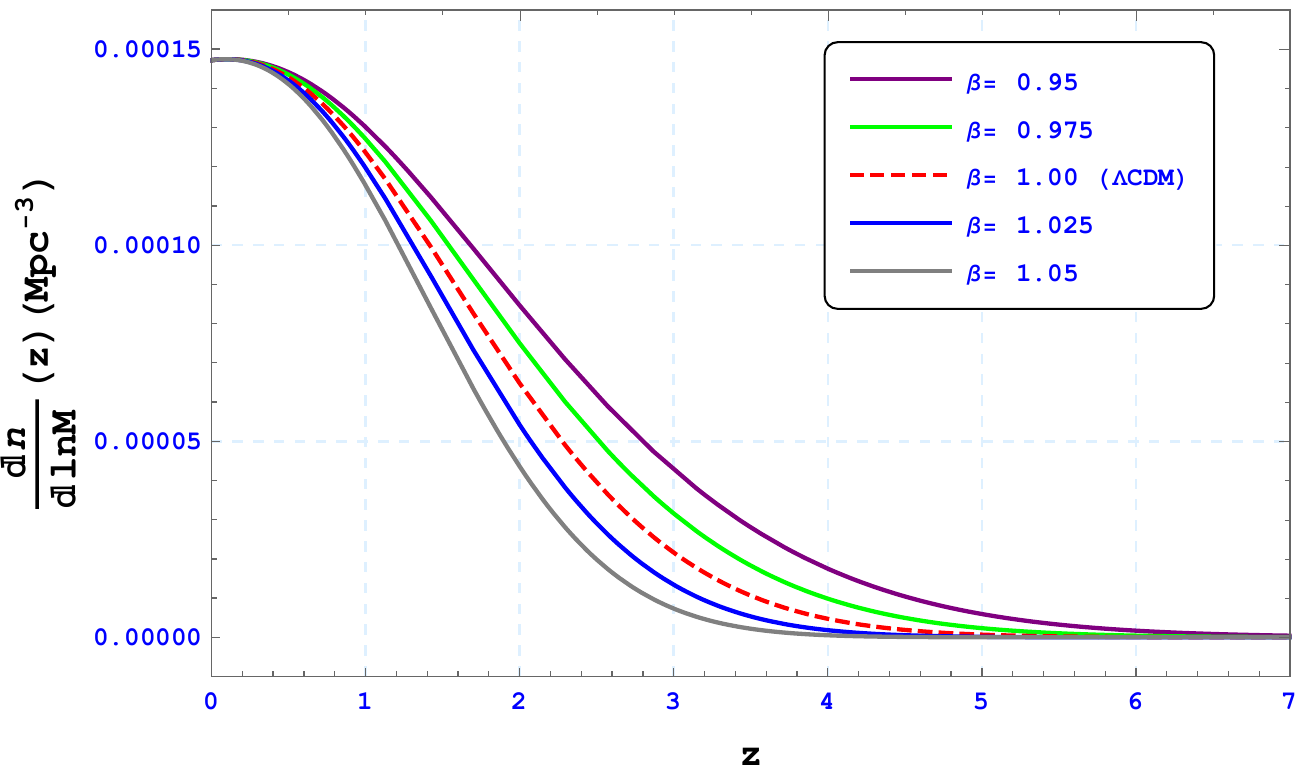}
    \caption{\small(Color online) Plot for Sheth, Mo, and Tormen--corrected differential halo mass function $\frac{dn}{d \ln M}(M,z)\big\rfloor_{SMT}$ versus redshift $z$ for mass $M=10^{13}\mathrm{h}^{-1}M_{\odot}$ with different values of TC model parameter $\beta$}
    \label{dnBYdlnmTC.pdf}
\end{figure}
The previously mentioned model has established itself as notably successful, as it shows agreement with the outcomes generated from simulation, especially within the moderate halo mass range. The emergence of precision cosmology, copupled with the rapid advancements in the resolution and scale of N-body simulations, has facilitated a solid prediction of numerical values of DHMF across multiple orders of magnitude \cite{watson2013halo,tinker2008toward,crocce2010simulating,courtin2011imprints,bond1991excursion,hellwing2016copernicus,pillepich2018first}. These findings indicate that original $\mathcal{F}_{PS}(M,z)$ significantly overestimates the quantity of low-mass halos, while underestimating the quantity of massive halo structures in the cluster mass range. In this regard, various revised models have been put forward in the literature (see Refs. \cite{tinker2008toward,crocce2010simulating,courtin2011imprints, sheth2001ellipsoidal,lukic2007halo,peacock1990alternatives,reed2013towards,watson2013halo}). While each of these alternative models have their own advantages and disadvantages \cite{Murray:2013qza}, it has been verified that most of the alternative DHMF models achieve similar accuracy \cite{gupta2022universality}. Thus, for the purpose of simplicity and conciseness, we select one specific model that accommodates an ellipsoidal peak shape, enables a moving barrier, and relaxes PS assumptions for our analytical predictions proposed by Sheth, Mo, and Tormen \cite{sheth2001ellipsoidal}
\begin{align}\label{F_SMT}
\mathcal{F}_{\mathrm{SMT}}(M, z) 
&= A \sqrt{\frac{2a}{\pi}}
\left[
1 + \left(
\frac{\sigma^2(M,z)}{a \delta_{cr}^2(z)}
\right)^p
\right]
\frac{\delta_{cr}(z)}{\sigma(M,z)}
\nonumber \\
&\quad \times
\exp\left(
-\frac{a \delta_{cr}^2(z)}
{2 \sigma^2(M,z)}
\right)\, ,
\end{align}
where $p=0.3$, $a=0.707$, and $A\approx 0.3222$ were derived from the simulation of DM halo formation. Theoretically, the critical density threshold $\delta_{cr}(z)$ is influenced by redshift and the model parameters of the governing cosmology. However, it has been noted that $\delta_{cr}(z)$ shows less sensitivity to their variations particularly after a few redshifts \cite{vianapedro1996cluster,herrera2017calculation,nunes2006structure,velten2014structure,abramo2007structure,farsi2022structure,farsi2023evolution,gupta2022universality,mukherjee2025spherical}. Furthermore, we also have verified that DHMF does not significantly change for a slight deviation from $\delta_{cr}=1.686$ \cite{padmanabhan1993structure}, which is associated with the Einstein-de Sitter (and the $\Lambda$CDM) Universe. Therefore, we shall apply the standard $\Lambda$CDM spherical collapse-based $\delta_{cr}=1.686$ to prevent extra model-dependent complexity for acquiring all the analytical predictions in the TC-modified cosmological framework with parameter $\beta$ that varies around the $\Lambda$CDM profile with $\beta =1$. The actual form of $\sigma(M,z)$ is typically approximated around a specific length $R_8 = 8h^{-1}\, Mpc$, corresponding to mass $M_8=5.95\times10^{14}\,\Omega_{m,0}\mathrm{h}^{-1}M_{\odot}$ \cite{herrera2017calculation} ($M_{\odot}$ represents the solar mass), and we adopt the fitting from Ref. \cite{vianapedro1996cluster}
\begin{equation}
    \sigma(M,z)=\sigma(M_8,0)\frac{\delta_m(z)}{\delta_m(z=0)}\left(\frac{M}{M_8}\right)^{-\frac{\gamma (M)}{3}}\, ,
\end{equation}
where 
\begin{equation}
    \gamma (M) = (0.3\,\mathrm{h}\,\Gamma+0.2)\left[2.92+\frac{1}{3}\log\left(\frac{M}{M_8}\right)\right]\,,
\end{equation}
and the shape parameter is provided by \cite{sugiyama1994cosmic,vianapedro1996cluster}
\begin{equation}
    \Gamma = \Omega_{m,0}\,\mathrm{h}\exp\left[-\Omega_{bar,0}\left(1+\frac{1}{\Omega_{m,0}}\right)\right]\,,
\end{equation}
with $\Omega_{bar,0} = 0.02230/\mathrm{h}^2$ \cite{herrera2017calculation} as the current baryon density. By using the expressions provided, Eq. (\ref{DHMF}) can be reformulated by applying SMT corrections as
\begin{align}\label{SMT_DHMF}
\frac{dn}{d \ln M}(M,z)\bigg\rfloor_{SMT} &=  \frac{\rho_{m,0}}{M}  \left| \frac{d \ln \sigma(M,z)}{d \ln M} \right| \nonumber \\&\quad \times
A \sqrt{\frac{2a}{\pi}}
\left[1 + \left(\frac{\sigma^2(M,z)}{a \delta_{cr}^2(z)}
\right)^p\right]\frac{\delta_{cr}(z)}{\sigma(M,z)}
\nonumber \\&\quad \times
\exp\left(-\frac{a \delta_{cr}^2(z)}{2 \sigma^2(M,z)}\right)\, .
\end{align}
We present the evolution of the SMT halo multiplicity function $\mathcal{F}_{\mathrm{SMT}}(M, z)$ and the SMT-corrected DHMF $\frac{dn}{d \ln M}(M,z)\big\rfloor_{SMT}$ against redshift in Fig. \ref{f_SMTTC.pdf} and Fig. \ref{dnBYdlnmTC.pdf}, respectively, for a standard mass $M=10^{13}\mathrm{h}^{-1}M_{\odot}$ with varying model parameter $\beta$. We note that both figures explicitly exhibit a similar trend. At both low and high redshifts, they resemble the $\Lambda$CDM profile. At lower redshifts, there is a rapid growth, indicating that halo abundance develops in the later stage of the Universe, and their value rise as $\beta$ decreases. Additionally, we notice that they are influenced not only by the background cosmology but also by the perturbative quantities such as the MDC and the critical overdensity.
\begin{figure}[htbp!]
    \centering
    \includegraphics[width=8cm,height=6cm]{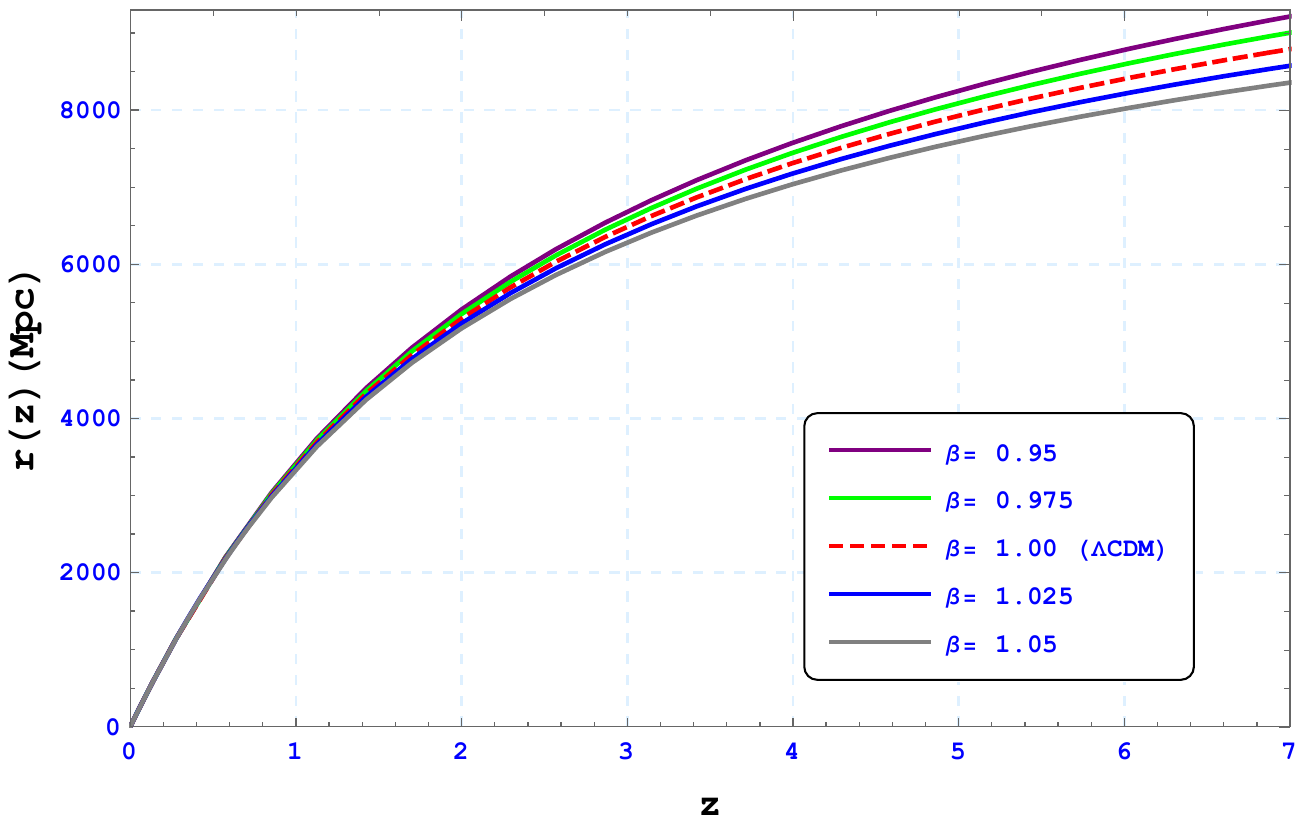}
    \caption{\small(Color online) Plot for comoving distance $r(z)$ versus redshift $z$  with different values of TC model parameter $\beta$}
    \label{rTC.pdf}
\end{figure}
\begin{figure}[htbp!]
    \centering
    \includegraphics[width=8cm,height=6cm]{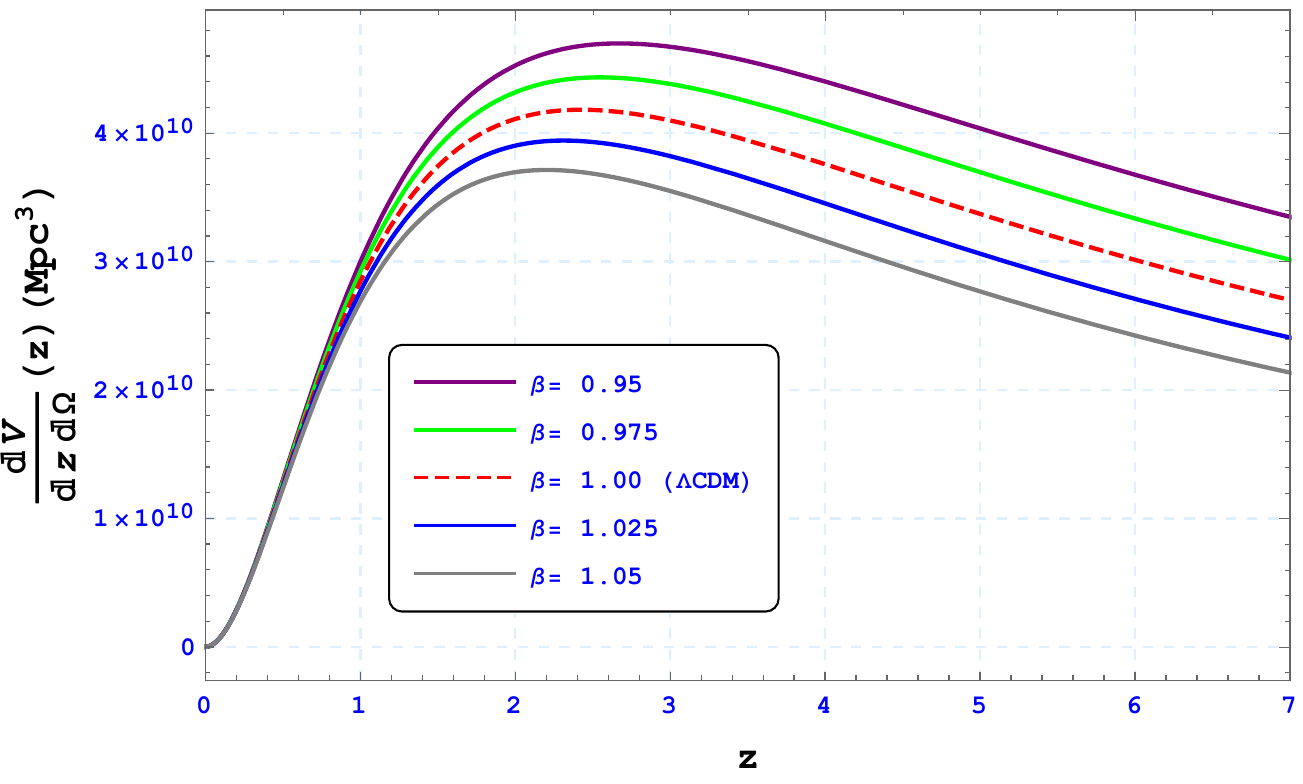}
    \caption{\small(Color online) Plot for comoving volume element $\frac{dV}{dz\,d\Omega}$ versus redshift $z$ with different values of TC model parameter $\beta$}
    \label{dVBYdzdOmegaTC.pdf}
\end{figure}
\begin{figure*}[htbp!]
   \raisebox{0.8in}{($a$)}\includegraphics[width=8cm,height=6cm]{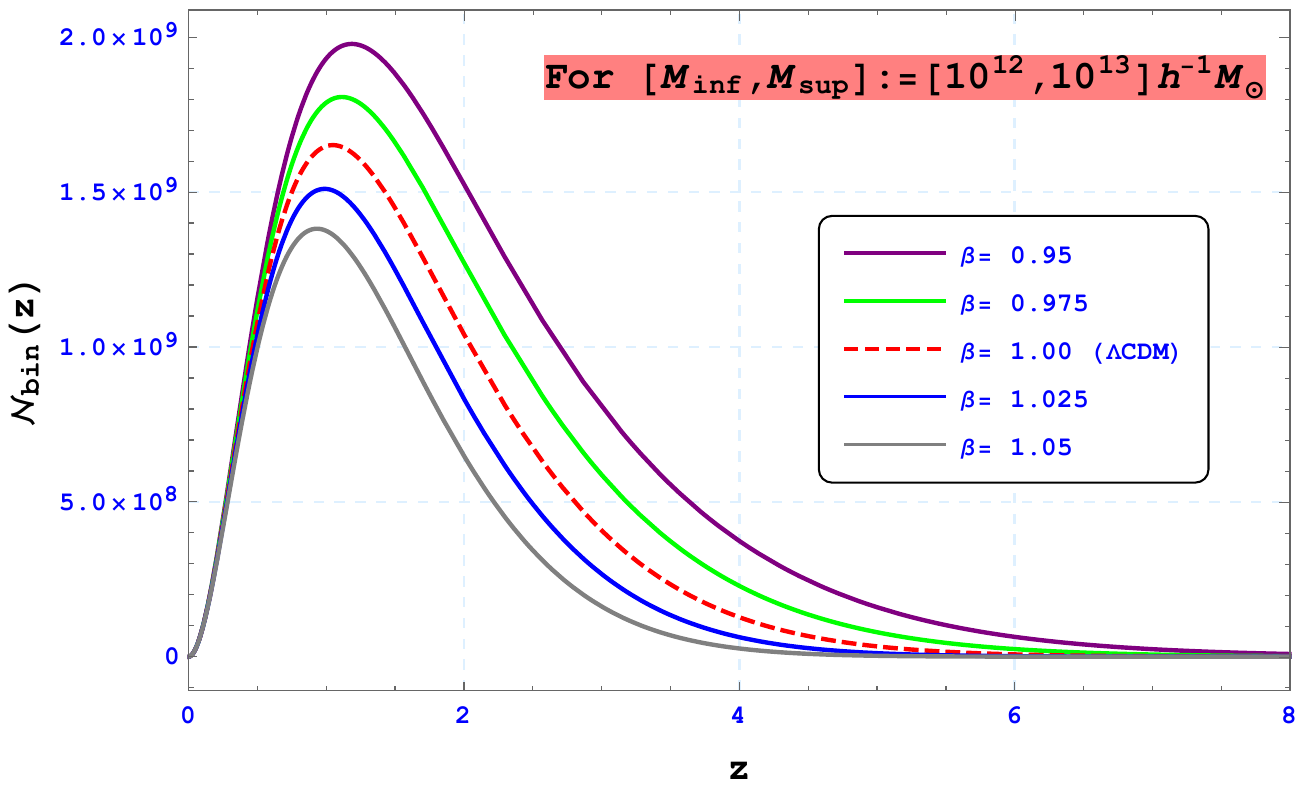}
   \hskip 0.03cm
  \raisebox{0.8in}{($b$)}\includegraphics[width=8cm,height=6cm]{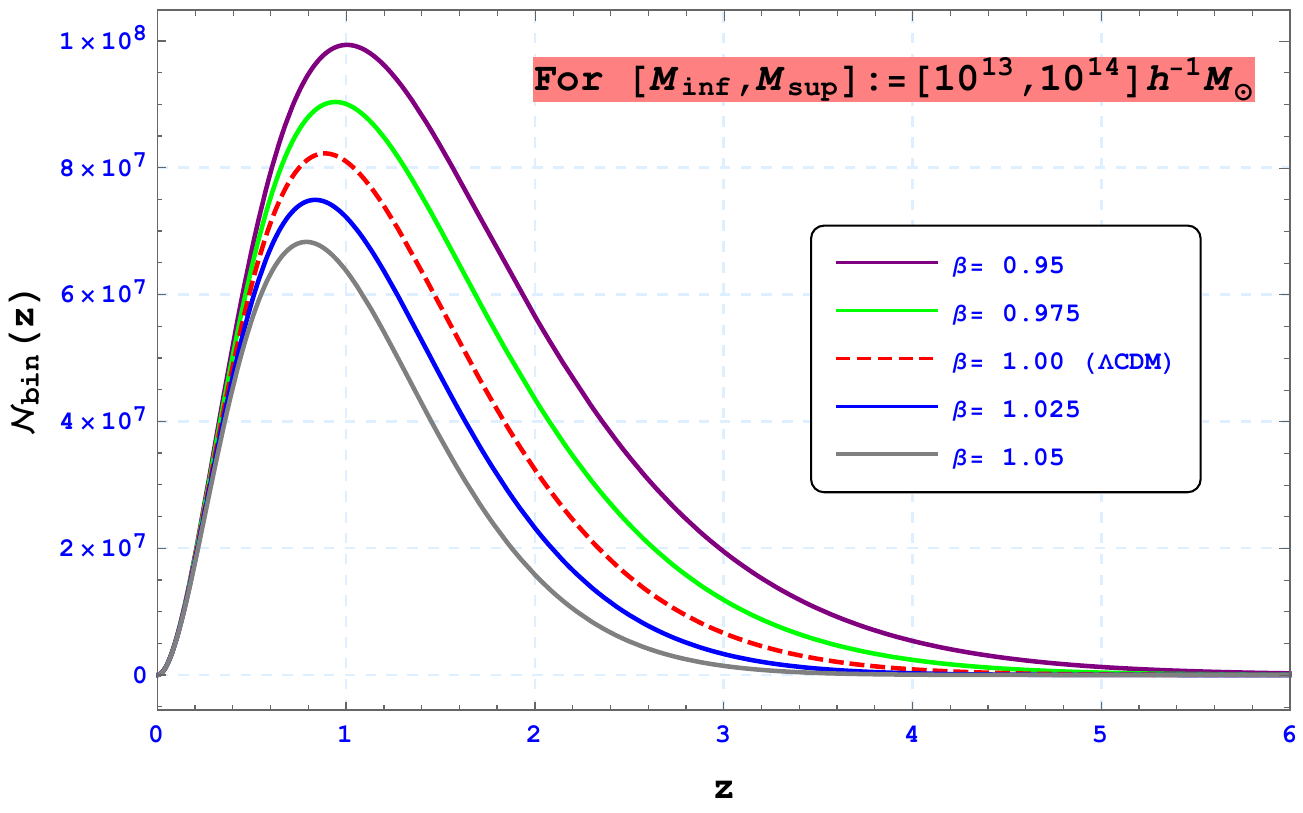}
  \caption{\small(Color online) Plots for halo number counts $\mathcal{N}_{bin}$ versus redshift $z$ for two specified mass bins with different values of TC model parameter $\beta$} 
 \label{N_bin plot TC}
\end{figure*}
\par Finally, we utilize the effective number of collapsed structures per unit of redshift  within a defined mass bin $M_{inf}<M<M_{sup}$ for the full sky survey of DM halo number count in the TC-modified cosmology \cite{nunes2006structure}
\begin{equation}\label{N_bin}
    \mathcal{N}_{bin} \equiv \frac{dN}{dz} \overset{\text{def}}{=} \int_{4\pi}d\Omega\int_{M_{inf}}^{M_{sup}}\frac{1}{M}\frac{dn}{d \ln M}\frac{dV}{dz\,d\Omega}dM\,,
\end{equation}
where
\begin{equation}
    \frac{dV}{dz\,d\Omega} \overset{\text{def}}{=} \frac{\,c\, r^2(z)}{H(z)}
\end{equation}
and
\begin{equation}
    r(z)\overset{\text{def}}{=}c\int_{0}^{z}H^{-1}(\tilde{z})d\tilde{z}
\end{equation}  
denotes the elementary comoving volume and the comoving distance, respectively. The redshift evolution of the comoving distance and the comoving volume element, together with the dependence of the model parameter $\beta$, is illustrated in Fig. \ref{rTC.pdf} and Fig. \ref{dVBYdzdOmegaTC.pdf}, respectively. It is observed that the comoving distance exhibits a monotonically increasing, concave-down growth pattern as we look further back in time. The elementary comoving volume initially rises from $z=0$, attains a maximum (around $1\lesssim z\lesssim3$, depending on the model parameter $\beta$), and then decreases at higher redshifts. Both quantities show an increase as the values of the model parameter $\beta$ decrease, and they align smoothly with the $\Lambda$CDM framework at the lower redshifts. Furthermore, these quantities are primarily determined by the background expansion, rather than the perturbative quantities. The halo number counts in mass bins in the TC-modified cosmology, $\mathcal{N}_{bin}\equiv \frac{dN}{dz}$, can be derived from Eq. (\ref{N_bin}), as shown in Fig. \ref{N_bin plot TC} for two distinct specified mass bins $[M_{inf}, M_{sup}]$ as $[10^{12}, 10^{13}]\mathrm{h}^{-1}M_\odot$ and $[10^{13}, 10^{14}]\mathrm{h}^{-1}M_\odot$. We observe that $\mathcal{N}_{bin}$ declines with an increase in the model parameter $\beta$, and a minute shift of the maximum redshift towards an earlier epoch for structure formation is observed for different values of the model parameter. At both extremes of the redshift spectrum, all the curves asymptotically converge with the $\Lambda$CDM profile. As all the curves reach their peak value around late times (around $0.5\lesssim z \lesssim 1.5$, depending on the model parameter), there may be an excessive number of structures forming during these later epochs. Moreover, a comparison of the two mass bins reveals a deficiency of heavier structures. These findings are consistent with the hierarchical model of large-scale structure formation, as supported by the works in Refs. \cite{abramo2007structure,liberato2006dark, farsi2022structure,farsi2023evolution,mondal2026cosmological}.
\section{Concluding remarks} \label{Discussion and Conclusions}
\par Implementing the nonadditive TC entropy in order to retain extensivity alongside the thermodynamics-gravity conjecture, we find that the expansion history of the Universe can be modified in comparison to the standard $\Lambda$CDM framework. In practical phenomenological terms, this can be envisioned as a potential extension to the standard $\Lambda$CDM cosmological framework. Specifically, we have used the Friedmann dynamical equations for the TC-modified cosmological model, incorporating matter and a cosmological constant $\Lambda$. By examining this new entropic effect on the linear growth of perturbations, we have investigated structure formation beyond the constraints of the $\Lambda$CDM paradigm, applying a top-hat SC formalism together with the SMT mass function. 
\par Similar to the standard cosmology, this modified cosmology indicates that the Universe experienced a transition from a deceleration to an acceleration phase at approximately a redshift $z_{\mathrm{tr}}\approx0.64$. Our investigation shows that the incorporation of the $\beta$-term in the TC-modified cosmology changes the epoch of transition. It is demonstrated that this model meets the criteria for the Universe to achieve thermodynamic equilibrium in the far future. The characteristics of the other cosmographic parameters, including jerk, snap, etc., confirm that the present Universe is accelerating. We introduce a new and long-standing diagnostic method to differentiate various cosmological models in the context of both flat and non-flat $\Lambda$CDM frameworks, and our findings reveal that the TC-modified cosmology ($\beta\ne 1$) passes all assessments, thus dismissing both the flat and non-flat $\Lambda$CDM models. In this context, by employing the top-hat SC approach and modified Friedmann dynamical equations, we derive and analytically solve the evolution equation of the linear density contrast. The linear MDC increases with decreasing redshift for all values of $\beta$. In this regard, we have examined growth rates $f(z)$ and $f(z)\sigma_8(z)$ for the TC-modified cosmology and discovered that there is a suppression of growth near lower redshifts, attributed to the DE dominance. The shifting of $f(z)\sigma_8(z)$ peaks to lower redshifts for higher values of $\beta$ suggests that the large-scale structures develop at later stages in the TC-modified cosmological scenario compared to the standard $\Lambda$CDM counterpart for $\beta>1$; conversely, they form earlier for $\beta <1$.  One of the prime objectives of our current analysis was to investigate the DM halo/cluster survey in the context of TC-modified cosmology. The changes introduced by the TC-modified cosmology affect the DM halo number counts, with smaller $\beta$ values resulting in higher halo abundances and vice versa. Additionally, it is also observed that the halo mass function begins to increase at lower redshifts, allowing halo structures to form at later epochs. Furthermore, we note that the halo number counts in mass bins, where the less massive structures are predominant, and they develop at later epochs. These findings perfectly align with the hierarchical model of large-scale structure formation. 
\par Our current research indicates that the TC-modified gravity scenario exhibits distinct signatures compared to the fiducial $\Lambda$CDM profile in all aspects. It raises an intriguing question of whether the TC entropic effect could simultaneously address the $H_0$ and $\sigma_8$ tensions and support its viability. To properly tackle these tensions, a thorough data analysis is required, exploiting various observational probes like  Supernovae Type Ia (SNe Ia), Cosmic Microwave Background Radiation (CMBR), Baryon Acoustic Oscillations (BAO), Cosmic Chronometers (CC), Hubble data, $f\sigma_8$ growth rate, and more. For halo abundance-related observations, there are initiatives relying on ongoing and upcoming observations, including the South Pole Telescope, eROSITA, etc. The South Pole Telescope survey, which is dependent on the detection of the Sunyaev–Zel'dovich effect, is expected to identify a substantial number of clusters and will be able to accurately measure the number counts \cite{carlstrom2002cosmology}. The evolution of the cluster mass function mirrors the linear growth of density fluctuations and can be employed to constrain the model parameters of modified gravity theories (See Ref. \cite{artis2024srg} for constraining model parameters in $f(\mathcal{R})$ gravity theory by analyzing the Hu-Sawicki parametrization together with the first Spectrum Roentgen Gamma (SRG)/eROSITA All-Sky Survey (eRASS$1$) cluster catalog in the western Galactic hemisphere, together with the overlapping data from KiloDegree Survey, Hyper Suprime-Cam, and Dark Energy Survey Year-$3$, for weak lensing mass calibration.). Undoubtedly, these observations can act as a robust cosmological probe to validate the feasibility of the TC-modified cosmological model and yield significant insights into it. Moreover, an investigation of nonlinear density perturbations for this modified model is also crucial and necessary regarding structure formation, as the observational evidence \cite{huterer2015growth} suggests that most of the structures originate from nonlinear evolution of overdensities in the dark age ($10 \lesssim z \lesssim 100$) \cite{miralda2003dark}. Nonetheless, these subjects lie outside the scope of our present study and will be addressed in future research.
\begin{acknowledgments}
SM expresses gratitude to the Government of West Bengal, India, for providing the State-funded Senior Research Fellowship (SRF).
\end{acknowledgments}
\section*{Data availability statement}
Data sharing does not apply to this article since no datasets were created or analyzed during this research.
\section*{Code availability statement}
Code/Software sharing does not apply to this article, as the present study did not include the generation or analysis of any code or software.
\section*{Conflict of interest}
The authors declare that they have no conflicts of interest to disclose.
\bibliography{reference}

\end{document}